\documentclass{aa} 
\usepackage{natbib}
\usepackage{graphicx}
\usepackage{txfonts}
\usepackage{times}
\usepackage{rotating}
\usepackage{float}
\usepackage{setspace}
\usepackage{lscape}
\usepackage{tabularx}
\usepackage{caption}
\usepackage[usenames]{color}
\usepackage[toc,page]{appendix}
\usepackage{soul}
\usepackage{amsmath}
\usepackage{hyperref}
\usepackage{siunitx}
\usepackage{threeparttable}
\usepackage{orcidlink}
\usepackage{lastpage}

\newcommand{\ppxf}{\textsc{pPXF}}

\newcommand{\HIIphot}{\textsc{\HIIphot}}

\newcommand{\msun}{\,M$_{\sun}$}

\def\deg{\mbox{$^{\circ}$}}

\begin{document}

\title{The Nuclear Star Cluster of M~74: a fossil record of the very early stages of a star-forming galaxy}

\subtitle{}

\authorrunning{}

\titlerunning{}

\newcommand{\MPIA}{\label{MPIA}Max-Planck-Institut f\"{u}r Astronomie, K\"{o}nigstuhl 17, D-69117, Heidelberg, Germany}


\newcommand{\IAC}{\label{IAC}Instituto de Astrof\'isica de Canarias, calle Vía L\'actea s/n, E-38205 La Laguna, Tenerife, Spain}

\newcommand{\ULL}{\label{ULL}Departamento de Astrof\'isica, Universidad de La Laguna, Avenida Astrof\'isico Francisco S\'anchez s/n, E-38206 La Laguna, Spain}

\newcommand{\UCA}{\label{UCA}Universit\'e C\^ote d'Azur, Observatoire de la C\^ote d'Azur, CNRS, Laboratoire Lagrange, 06000, Nice, France}

\newcommand{\UWyoming}{\label{UWyoming}Department of Physics and Astronomy, University of Wyoming, Laramie, WY 82071, USA}

\newcommand{\stromlo}{\label{stromlo}Research School of Astronomy and Astrophysics, Australian National University, Mt Stromlo Observatory, Weston Creek, ACT 2611, Australia}

\newcommand{\ITA}{\label{ITA}Universit\"{a}t Heidelberg, Zentrum f\"{u}r Astronomie, Institut f\"{u}r Theoretische Astrophysik, Albert-Ueberle-Str 2, D-69120 Heidelberg, Germany}

\newcommand{\IWR}{\label{IWR}Universit\"{a}t Heidelberg, Interdisziplin\"{a}res Zentrum f\"{u}r Wissenschaftliches Rechnen, Im Neuenheimer Feld 205, D-69120 Heidelberg, Germany}

\newcommand{\OAN}{\label{OAN}Observatorio Astron\'{o}mico Nacional (IGN), C/Alfonso XII, 3, E-28014 Madrid, Spain}

\newcommand{\JBCfA}{\label{JBCfA}UK ALMA Regional Centre Node, Jodrell Bank Centre for Astrophysics, Department of Physics and Astronomy, The University of Manchester, Oxford Road, Manchester M13 9PL, UK}

\newcommand{\ADOhio}{\label{ADOhio}Astronomy Department, The Ohio State University, Columbus, Ohio, USA
}

\newcommand{\CCOhio}{\label{CCOhio}Center for Cosmology and Astro-Particle Physics, The Ohio State University, Columbus, Ohio, USA
}

\author{Francesca Pinna \orcidlink{0000-0001-5965-3530}\inst{\ref{IAC},\ref{ULL},\ref{MPIA}}\thanks{\email{francesca.pinna@iac.es}}
          \and
          Nils Hoyer \inst{\ref{MPIA}}
          \and
          Jairo M\'endez Abreu \inst{\ref{ULL}, \ref{IAC}}
          \and 
          Adriana de Lorenzo-C\'aceres Rodriguez \inst{\ref{ULL},\ref{IAC}}
          \and 
          Nadine Neumayer \inst{\ref{MPIA}}
          \and
          Médéric Boquien \inst{\ref{UCA}}
          \and 
          Salvador Cardona Barrero \inst{\ref{ULL},\ref{IAC}}
          \and
          Daniel~A.~Dale \inst{\ref{UWyoming}}
          \and
          Ivan S. Gerasimov\orcidlink{0000-0001-7113-8152}\inst{\ref{UCA}}
          \and
          Kathryn~Grasha 
          \orcidlink{0000-0002-3247-5321}
          \inst{\ref{stromlo}}
           \and 
           Ralf S.\ Klessen \orcidlink{0000-0002-0560-3172}\inst{\ref{ITA},\ref{IWR}} 
          \and
          Carlos Marrero de la Rosa \inst{\ref{ULL}, \ref{IAC}}
          \and
          Miguel Querejeta \orcidlink{0000-0002-0472-1011}\inst{\ref{OAN}} 
          \and
          Thomas G. Williams \orcidlink{0000-0002-0012-2142}\inst{\ref{JBCfA}} 
          \and
          Smita Mathur \orcidlink{0000-0002-4822-3559}\inst{
          \ref{ADOhio}, \ref{CCOhio}
          }
          \and
          Eva Schinnerer \inst{\ref{MPIA}}
          }

\institute{\IAC \and \ULL \and \MPIA \and \UCA \and \UWyoming \and \stromlo \and \ITA \and \IWR \and \OAN \and \JBCfA \and \ADOhio \and \CCOhio
}

\date{\today}

\abstract{
Nuclear star clusters (NSC) are dense and compact stellar systems, of sizes of few parsecs, located at galactic centers. 
Their properties and formation mechanisms seem to be tightly linked to the evolution of the host galaxy, with potentially different formation channels for late- and early-type galaxies (respectively, LTGs and ETGs). While most observations target ETGs, here we focus on the NSC in M~74 (NGC~628), a relatively massive, gas-rich and star-forming spiral galaxy, part of the PHANGS survey. We analyzed the central arcmin of the PHANGS-MUSE mosaic, in which the NSC is not spatially resolved. 
We analyzed the NSC stellar populations in a point spread function (PSF) aperture, and compared it to the host galaxy. Within the PSF size, the NSC is contaminated by the host-galaxy light. We performed a two-dimensional spectro-photometric decomposition of the MUSE cube, employing a modified version of the C2D code, to disentangle the NSC from its host. 
This method provided different data cubes for the NSC and the host galaxy, allowing for both their comparison in a PSF aperture, and the spatially resolved analysis of the host. 
Our results show a very old and metal-poor NSC, in contrast to the surrounding regions. While similar properties were found in NSCs hosted by galaxies of different masses and/or morphological types from M~74, they are somewhat unexpected for a relatively massive star-forming spiral galaxy. The spatially resolved stellar populations of the host galaxy display much younger (light-weighted) ages and higher metallicities, especially in the central region (${\sim}500$~pc) surrounding the NSC. This suggests that this NSC formed a long time ago, and evolved passively until today, without any further growth. No significant amounts of gas would have reached the very central region in the last 8~Gyr.
}


\maketitle

\section{Introduction}
\label{sec:intro}

In the central regions of galaxies, massive black holes often coexist with dense stellar systems known as nuclear star clusters (NSCs). NSCs have sizes of a few parsecs and masses between $10^4$ and a few times $10^8$\msun \citep[e.g.,][]{Boeker2002,Boeker2004, Walcher2005, Cote2006, Georgiev2016, SanchezJanssen2019, Hoyer2023a}. Their properties trace the dynamical state and evolution of the central regions of galaxies and are key to revealing not only their formation and growth but also the mechanisms funneling stars and gas to the center, leading to the seeding and growth of central black holes (see \citealt{Neumayer2020} for a review). 
NSCs can form via two main channels: star-cluster inspiral to the center due to dynamical friction \citep[e.g.,][]{Tremaine1975, Capuzzo-Dolcetta1993, Oh2000, Lotz2001, Arcasedda2014} and in-situ star formation after gas inflow \citep[e.g.,][]{Silk1987, Mihos1994, Milosavljevic2004, Bekki2006, Bekki2007}. 
These two formation channels are not mutually exclusive, as migrating star clusters can be gas-rich \citep{Guillard2016}, and combinations of in-situ formation and cluster merging have often been necessary to explain the observed properties \citep{Fahrion2019, Fahrion2021, Fahrion2022a, Fahrion2022b, Fahrion2024}. 

More than 60\% of galaxies with stellar masses between $10^8$ and $10^{10}$\msun{} host an NSC, with a peak nucleation fraction (${\sim}90$\%) near $10^{9}$\msun{} \citep{SanchezJanssen2019, Hoyer2021}. This fraction drops drastically for lower masses. For higher masses, a similar drop is observed for early-type galaxies (ETGs) hosting NSCs, while massive late-type galaxies (LTGs) seem to maintain a similar high fraction \citep{Neumayer2020,Ashok2023}. This trend, together with NSC properties such as metallicity, suggests a transition in the dominant NSC formation channel at about $10^{9}$\msun\, \citep[e.g.][]{Lyu2025}: many low-mass galaxies host metal-poor NSCs, which have similar properties to globular clusters and are traditionally associated with cluster inspiral \citep[e.g.,][]{Fahrion2020}. NSCs are systematically more metal-rich at higher galaxy masses, showing significant contributions from younger stellar populations, often ongoing or recent star formation (see, e.g., Fig.~9 in the review from \citealt{Neumayer2020}, and \citealt{Fahrion2021, Fahrion2022a, Fahrion2022b}). 
However, metal-rich NSCs may also form from the inspiral of young massive star clusters, complicating the interpretation of observations \citep{Paudel2020, Fahrion2024}. 

Most evidence of the trends described above comes from NSCs hosted by ETGs, while it needs to be further investigated in LTGs. 
Several kinematic and stellar-population studies revealed morphological dependencies. 
\citet{Pinna2021} analyzed the spatially resolved kinematics of eleven NSCs. While rotation is ubiquitous in the full sample, NSCs in LTGs are typically rotation-dominated and associated with central gas inflow, whereas those in ETGs show slower rotation, more complexity and signatures of past mergers (especially at lower NSC masses). 
NSCs in very LTGs (Scd or later types) are on average young, 
but made up of stellar populations of different ages, with prolonged and recurrent star formation until the present day \citep{Walcher2005,Walcher2006, Rossa2006, Kacharov2018}. 
NSCs in LTGs also showed a diversity in their ages and metallicities \citep{Kacharov2018}. 
Despite the scatter, NSCs follow an average trend being generally older in earlier-type spirals (${\sim}1-2$~Gyr) and younger, less massive, and more metal-poor in the latest types (about a couple of hundred Myr, \citealt{Rossa2006, Walcher2006}). 
Combined trends with galaxy mass and morphological types support a link between NSC formation and the host-galaxy evolution. 

While simulations have reproduced aspects of both formation channels, systematic theoretical studies across galaxy types are still lacking.
On the one hand, individual parsec-scale resolution $N$-body simulations were essential to show that star cluster inspiral can easily lead to the observed properties of NSCs \citep[e.g.,][]{Mastrobuono2014, Arcasedda2014, ArcaSedda2020, Mastrobuono2023,Mastrobuono2025}. On the other hand, other simulations have shown that in-situ formation is needed to explain the complexity in the kinematics and stellar population properties of stars in the center, e.g., of the Milky Way \citep[e.g.,][]{Mapelli2012a, Mastrobuono2019}. 
However, in numerical simulations, larger galaxy samples and a cosmological context are needed to assess the impact of internal and external galaxy-evolution processes in the formation of nuclear structures, while the study of NSCs requires very high spatial and mass resolution, posing a serious computational challenge. 
Cosmological simulations of galaxies at redshift $z{\sim}1.5$ by \citet{Brown2018} showed the presence of central clusters (made up of only few stellar particles), with extended star formation histories (SFHs) leading to large spreads in ages and metallicities, and global higher metallicities than their host galaxy. 
A recent study using the Engineering Dwarfs at Galaxy Formation's Edge (EDGE) simulations \citep{Agertz2020} traced back the emergence of an NSC in four models of low-mass dwarf galaxies (with stellar masses 
between $10^6$ and $10^7$\msun, \citealt{Gray2025}). The NSC forms at redshift $z{\sim}2$ during a starburst triggered by a major merger, but also contains a previously formed population of stars. 

Despite this progress, NSC formation remains unclear, and the relative importance of different growth mechanisms is still debated. Spatially resolving NSCs with spectroscopic observations is currently feasible only in nearby bright galaxies.
Although integral-field spectroscopy (IFS) enables comparisons between the NSC and its host galaxy, in most cases, NSCs are still unresolved, and we are limited to their point-spread-function (PSF) integrated spectra \citep[e.g.,][]{Fahrion2021, Fahrion2022b}. Yet, due to the small size of these objects, these studies are challenging and IFS samples are mostly restricted to ETGs (see above), mainly because lower surface brightness in LTGs requires long integration times. While NSCs in some dwarf LTGs were recently covered \citep{Fahrion2022b}, additional work is needed on massive spirals.

Accurate measurements further require disentangling NSC light from that of the host galaxy, to limit the contamination from other internal structures overlapped in the line of sight (LOS). 
This problem has been approached in different ways in the past. 
\citet{Johnston2020} used \textsc{BUDDI} (Bulge–Disc Decomposition of IFU data, \citealt{Johnston2017}) to model the light of the NSC and the host galaxy in two dimensions, for each wavelength, and separate one spectrum for each component. They modeled the NSC as a PSF profile and the host galaxy as a combination of one to four S\'ersic profiles. This ensured a minimal contamination of the NSC spectrum from the host galaxy. 
\citet{Fahrion2021} simply obtained the NSC spectrum by subtracting a representative spectrum of the host galaxy from a central PSF-weighted spectrum. The host-galaxy spectrum was obtained by multiplying a spectrum extracted from an elliptical annulus around the central region (at $160-200$pc from the center) by a scaling factor, calculated by modeling images for the NSC (as a PSF) and the host galaxy (the latter, modeled as a combination of a bulge and a disk). 

In this work, we present a new version of the spectro-photometric decomposition method \textsc{C2D} that enables the extraction of the stellar populations of the NSC while mapping its surroundings. \textsc{C2D} \citep{MendezAbreu2019} separates an IFS data cube into multiple morphological components, delivering a data cube for each one of them.
It uses a very similar approach to BUDDI (see above), with the difference that the latter delivers one spectrum for each morphological component, instead of a full, separated data cube as C2D does. This allows us to spatially resolve stellar populations in the NSC surroundings, helping understand the NSC evolution in the context of its host galaxy. 
We applied this method to the central region of the massive spiral galaxy M~74 (NGC~628) and analyzed its stellar populations. 
The paper is structured as follows. In Sect.~\ref{sec:obs}, we describe the data set. Section~\ref{sec:galaxy} summarizes the known properties of the galaxy M~74 and its NSC. In Sect.~\ref{sec:method}, we describe the methods used for the spectro-photometric decomposition of the NSC from the host galaxy, and for the stellar-population analysis. The analysis of the non-decomposed IFS data cube is presented in Appendix~\ref{app:res:orig_integr}. Section~\ref{sec:results} describes our results, which are discussed in Sect.~\ref{sec:disc}. Our conclusions are outlined in Sect.~\ref{sec:concl}.  


\section{Observations}
\label{sec:obs}
PHANGS (Physics at High Angular resolution in Nearby GalaxieS)\footnote{\url{ http://www.phangs.org}} is a multiwavelength survey aimed at tracing the small-scale physics of star formation through different phases of the interstellar medium. It covers at high spatial resolution a total sample of 74 nearby (closer than 20~Mpc), relatively face-on (with lower inclinations than 75\deg), massive and star-forming galaxies - with stellar masses log($M_*/$\msun)~$>9.75$ and specific star-formation rates log(sSFR yr$^{-1}) \gtrsim -11$. PHANGS includes large programs from the Atacama Large Millimeter Array (ALMA), the Hubble Space Telescope (HST), the James Webb Space Telescope (JWST), the Very Large Telescope (VLT) and Chandra, plus other ancillary data. 
In this work, we use PHANGS IFS data from the Multi Unit Spectroscopic Explorer (MUSE, \citealt{Bacon2010,Bacon2014}), mounted at the UT~4 of the VLT, in the Wide Field Mode (WFM). MUSE WFM has a field of view (FoV) of 1~arcmin$^2$ sampled at 0.2~arcsec~pixel$^{-1}$. It covers a wavelength range between 4800 and 9300~\AA, sampled at 1.25~\AA~pixel$^{-1}$ at a nominal spectral resolution of 2.5~\AA~(full-width-half-maximum, {FWHM}) at 7000~\AA. 
The PHANGS-MUSE survey \citep{Emsellem2022} focuses on 19 galaxies previously observed with ALMA, providing mosaics of up to 15 pointings for each galaxy. 

M~74 was observed with 12 pointings covering the full disk (\citealt{Kreckel2016, Kreckel2017, Kreckel2018}; Program IDs: 094.C-0623, 095.C-0473, 098.C-0484; P.I.s: K. Kreckel and G. Blanc). An image reconstructed from the full PHANGS-MUSE mosaic is shown in Fig.~9 of \citeauthor{Emsellem2022} (\citeyear{Emsellem2022}, see also Fig.~1 in \citealt{Kreckel2018}). 
Each pointing was observed with three exposures of 845 or 990~seconds (depending on the pointing), rotated 90\deg{} with respect to each other. 
Sky exposures were taken in between the object exposures. 
The data reduction was carried out using the dedicated implementation of the MUSE data reduction recipes \citep{Weilbacher2020} into the wrapper \texttt{pymusepipe}\footnote{\url{https://pypi.org/project/pymusepipe/}}. 
The data reduction process included corrections from bias, flat field, wavelength calibration, line-spread-function derivation, geometry and astrometry corrections, illumination correction through twilight sky ﬂats, sky subtraction, alignment and mosaicking. 
The full data reduction process was detailed by \citet{Emsellem2022}. 
The reduced mosaic is publicly available in the ESO Archive\footnote{\url{https://archive.eso.org/scienceportal/home?data_collection=PHANGS}}.

Because individual pointings were observed at different times and thus with different atmospheric conditions, the PSF varied between each other and was modeled for the PHANGS-MUSE survey for different pointings \citep{Emsellem2022}, using a circular Moffat function, which described it well \citep{Fusco2020}. 
The Moffat FWHM and power index were measured for each pointing using a cross-convolution method (by comparing with a reference image) and were assumed constant within each MUSE FoV \citep{Emsellem2022}. 
In this work, we used a PSF FWHM of 3 MUSE spatial pixels (3~spaxels, 0.6~arcsec) for the NSC decomposition, which approximates the average FWHM measured in the four pointings contributing to the central arcmin. The power index estimate was 2.8 for all pointings. 
The MUSE PSF also varies with wavelength, and it was measured by \citet{Emsellem2022} at a reference wavelength of 6483.58~\AA. 
We tested that a variation of up to $\pm 0.2$~arcsec (one pixel) in the FWHM did not affect the results (Sect.~\ref{sub:decomp}). 

\begin{figure}
\centering
\resizebox{0.5\textwidth}{!}
{\includegraphics[scale=1.]{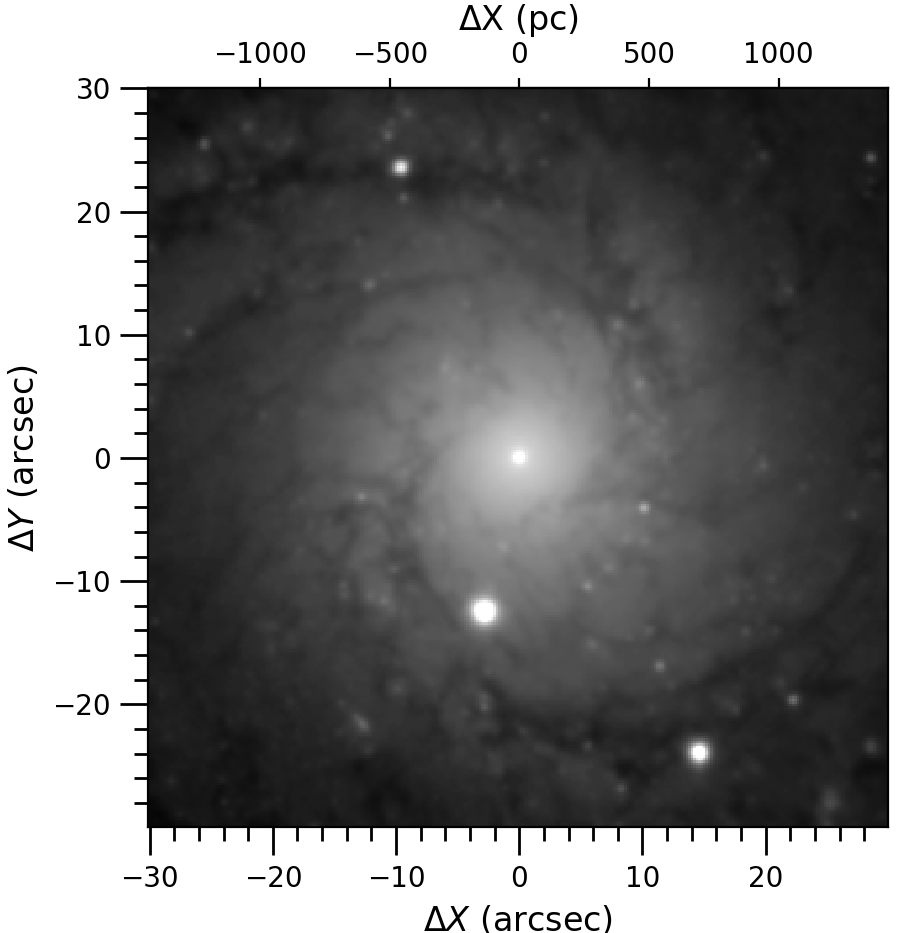}}
\caption{White image obtained from the cropped MUSE data cube of the central arcmin, combining the light in the full MUSE wavelength range. 
}
\label{fig:image}
\end{figure}
\section{M~74 (NGC~628)}
\label{sec:galaxy}

M~74 (NGC~628, or the ''Phantom Galaxy'') is a grand-design non-barred spiral galaxy classified as an SA(s)c \citep{Buta2015}. It lies at a distance of 9.84~Mpc and has a stellar mass of $2.2\times 10^{10}$\msun\,\citep{Leroy2021}. 
Its nearly face-on inclination (${\sim}8.9 \deg$,  \citealt{Lang2020}) is optimal to disentangle properties of the stellar nucleus from the host galaxy. 
Morphologically, M~74 consists of a faint bulge and a dominating disk, with no obvious bar \citep{Salo2015, Querejeta2021, Stuber2023}, though remnants of a past bar and a possible nuclear ring have been debated \citep[e.g.,][]{SanchezBlazquez2014b}. 

M~74 is a star-forming galaxy, with a star-formation rate of $\log_{10} \, (\textrm{SFR} / [\textrm{M}_{\odot} / \textrm{yr}]) = 0.24$, which places it above the star-formation main sequence of local galaxies \citep{Leroy2021, Kim2023}. 
Its star formation and interstellar medium have been extensively studied at high spatial resolution by the PHANGS team using a combination of MUSE, ALMA, HST, and JWST. 
Massive star-forming regions are observed both in the spiral arm and interarm regions \citep{Kreckel2016, Williams2022}. On large scales, star formation mostly follows the molecular-gas structure, while offsets at the scale of molecular clouds imply longer gas depletion times than in the Milky Way \citep{Kreckel2018}. 
HST and JWST images revealed with unprecedented detail a complex web of dust and gas filaments, ranging from compact molecular-cloud-sized structures to faint arm-tracing features \citep{Thilker2023}, along with stellar-feedback-driven bubbles and voids \citep{Barnes2023, Watkins2023}. A remarkable cavity of a size of about $200 \times 400$~pc$^2$, with no ionized or molecular gas, and no dust, was found in the center of the galaxy \citep{Kreckel2018, Herrera2020, Leroy2021, 
Stuber2023, Hoyer2023b}. 

M~74's spatially resolved stellar populations show radial trends. 
At large scale, \citet{SanchezBlazquez2014b} 
found predominantly old ages (${\sim} 10$~Gyr) beyond ${\sim} 6$~kpc, with a negative gradient consistent with an inside-out disk formation. 
The SFH reveals a dominant ${\sim} 13$~Gyr component, plus a minor recent ${\sim} 1$~Gyr component. 
The metallicities are slightly subsolar with a relatively flat gradient within ${\sim}3$~kpc, which becomes negative outwards. 
A more recent PHANGS-MUSE analysis \citep{Pessa2023} confirmed these trends at ${\sim}100$~pc resolution, although with average light-weighted and mass-weighted ages in the central kpc, respectively, of about 2 and 9~Gyr, and the youngest stellar populations along the spiral arms. 
Metallicity gradients are negative in the central kpc and flatter (slightly positive) in the outskirts, again supporting an inside-out assembly scenario. 
JWST resolved-star analyses further confirmed the negative metallicity gradient \citep{Ck2025}.
Finally, the star-cluster analysis from HST data has shown a distribution of ages ranging from a few million years, for cluster masses between $10^3$ and $10^5$\msun, to more than 10 billion years above $10^5$\msun (up to $10^7$\msun, \citealt{Whitmore2023}).

M~74 is the brightest member of the M~74 Group, in which dwarf satellite galaxies (with stellar masses $\rm 10^4\, M_\sun < M_* < 10^6$\msun) identified by \citet{Davis2021} show signs of recent quenching. 
M~74's HI disk shows a complex structure with a warp, high velocity complexes, and an asymmetric outer tail \citep{Kamphuis1992}. These features might be the result of one or more past accretion events. 

\subsection{The nuclear star cluster (NSC) of M~74} \label{sub:galaxy_nsc}

M~74's NSC was first identified with HST by \citet{Ganda2009} and later morphologically analyzed by \citet{Georgiev2014} and \citet{Georgiev2016}. 
A detailed study by \citet{Hoyer2023b}, combining JWST and HST imaging, confirmed these results and revealed additional structural complexity. 
They derived a NSC effective radius of $R_{\rm eff} {\sim}5$pc, an ellipticity ($\epsilon$) of 0.05, a S\'ersic index between 2 and 3, using ultraviolet, optical and near-infrared band filters. 
JWST Mid-Infrared Instrument (MIRI) data gave a larger size of $R_{\rm eff} {\sim}12$~pc, higher ellipticity ($\epsilon {\sim}0.4$), and a S\'ersic profile close to exponential ($n {\sim}1.5$). This suggests multiple components or wavelength-dependent substructures. 
\citet{Hoyer2023b} discussed different explanations for the mid-infrared emission, including the contribution from a weak AGN, dust released by asymptotic-giant-branch (AGB) stars, an infalling star cluster, or a circumnuclear disk. 

Spectral Energy Distribution (SED) fitting across ten bands from the ultraviolet to the near infrared yielded an old age of $8 \pm 3$~Gyr and a subsolar metallicity $Z=0.012 \pm 0.006$. This suggested that the NSC was not involved in relatively recent star formation (within few Gyr). 
The inclusion of MIRI data degraded the fit. 
The PHANGS-JWST observations showed that the NSC resides in the central cavity devoid of gas and dust, contrasting with the dense and complex web of filaments of the galaxy (see Sect.~\ref{sec:galaxy}). 
The old ages of the NSC suggest that the cavity has persisted for several Gyr at least. 
It may have formed just by gas depletion combined with some mechanism preventing gas from inflowing to the center, or by the evacuation of the previously present gas. 
\citet{Hoyer2023b} obtained a NSC stellar mass of ${\sim}1.26 \times 10^{7}$\msun. 
While this is a relatively massive NSC compared to the average \citep[e.g.,][]{Fahrion2022b}, it is less massive than NSCs in galaxies of similar masses as M~74 \citep{Hoyer2023b}. This is consistent with limited subsequent in-situ growth due to the gas-free environment. 

Archival Chandra data reveal an X-ray source associated with the NSC \citep{Lehmer2024}, whose X-ray spectrum 
is consistent with one or more low-mass X-ray binaries (LMXBs) in which a low-mass star accretes on a stellar-mass black hole or a neutron star. 
The presence of LMXBs is a signature of old stellar populations and they are usually found in elliptical galaxies and globular clusters. 
While an AGN origin cannot be ruled out, it is disfavored given the association of the X-ray source with the NSC. 


\section{Methods} \label{sec:method}

\begin{figure*}
\centering
\resizebox{1.\textwidth}{!}
{\includegraphics[scale=1.]{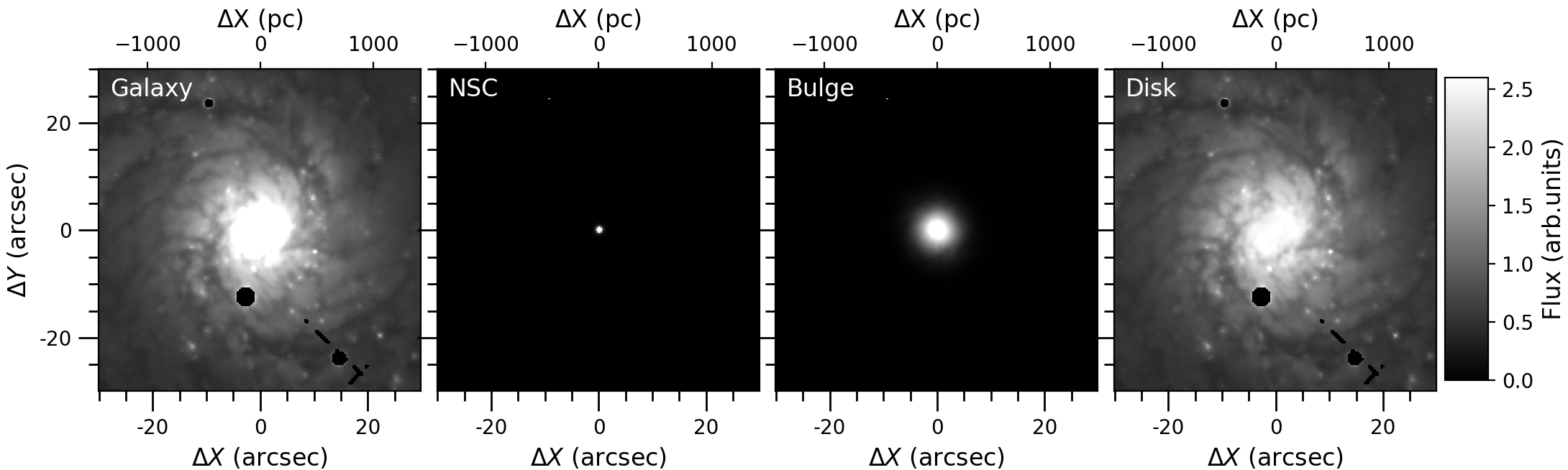}}
\caption{From left to right: full kinematic-corrected data cube (given to C2D as an input), NSC, bulge and disk, the three of them resulting from the C2D spectrophotometric decomposition. Masked regions are plotted in black. 
}
\label{fig:datacubes}
\end{figure*}

The goal of this paper is to analyze the stellar populations of the NSC in M~74 in the context of the host galaxy, while we are not interested in the kinematic properties (they would be unresolved for the NSC). We first disentangled the NSC light from its host using a spectro-photometric decomposition method, yielding separate data cubes for each component (Sect.~\ref{sub:decomp}). For the stellar population analysis, we used two overall approaches. One consisted of fitting spectra integrated over a PSF aperture (Sect.~\ref{sub:c2d_m74}), separately for the NSC and the host galaxy (one each, Fig.~\ref{fig:spectra}). Secondly, we fitted the spectra of each spaxel in the host-galaxy data cube, allowing the spatially resolved analysis of its stellar populations. All the details about the spectral fitting process are given in Sect.~\ref{sub:spectr_fit}.
\subsection{Spectrophotometric decomposition of the NSC: C2D} \label{sub:decomp}

As mentioned in Sect.~\ref{sec:intro}, disentangling the NSC light from the other galaxy components, overlapped in the central region along the LOS, is essential to analyze the properties of the NSC more accurately. 
For this purpose, we used \textsc{C2D}, a multicomponent spectrophotometric decomposition code \citep{MendezAbreu2019}. 
C2D creates one quasi-monochromatic image for each spectral resolution element, and then performs a multicomponent photometric decomposition of those images. 
The photometric decomposition of the galaxy surface brightness distribution in each image is done in two dimensions with the code GAlaxy Surface Photometry 2 Dimensional
Decomposition (\textsc{GASP2D}, \citealt{MendezAbreu2008}). 
\textsc{GASP2D} adopts a Levenberg-Marquardt algorithm to estimate the structural parameters that fit the two-dimensional surface brightness of the galaxy. In the first version, it was optimized to fit a bulge and a disk \citep{MendezAbreu2008}, while it was later adapted to fit also bars \citep{MendezAbreu2014}. In this work, we adapted \textsc{GASP2D} to fit an unresolved NSC, in addition to a bulge and a disk (Sect.~\ref{sub:c2d_m74}). 

The fit for different wavelengths with C2D is done in different steps. First, the wavelength range is divided into ten wide spectral bins. An image of a relatively high S/N is created by integrating all spectral pixels in each wavelength bin. The first morphological fit is done for each one of these images, using some initial conditions for the free parameters. These initial conditions are taken from the interpolation of some constraints given in the three $g$, $r$ and $i$ bands by the user. 
Then, the parameters fitted for the ten bins are interpolated for all wavelengths of the data cube, and an image for each wavelength is fitted, now keeping as a free parameter only the intensity of each profile. The final step is the computation of the fraction of light of each component in each spatial and spectral pixel. This allows to reconstruct an entire data cube for each of the three components, preserving the spatial information. 
The details of the setup used here are given in Sect.~\ref{sub:c2d_m74}. 

\subsubsection{Preparation of the data cube of the central region of the galaxy} \label{sub:preparation}

In this section, we describe all the steps needed to prepare the data for the C2D spectrophotometric decomposition. We used, for the full analysis presented in this paper, the kinematic-corrected data cube resulting from all the steps described here. 

\begin{enumerate}
\item \textit{Cropping the mosaic.}
Since this work focuses on the central region of the galaxy, we first cropped the PHANGS-MUSE mosaic to the central arcmin. 
This ensured a successful NSC decomposition, avoiding the need to handle the full mosaic, which is computationally expensive. This also avoided offsets in the stellar population parameters between different MUSE pointings, which were detected in PHANGS-MUSE large mosaics due to biases in the sky subtraction \citep{Emsellem2022}. 
An image obtained from the cropped data cube of the central arcmin, using the full MUSE wavelength range, is shown in Fig.~\ref{fig:image}. We also created a mask to exclude bright foreground stars from the analysis described below (see, e.g., Fig.~\ref{fig:datacubes} and \ref{fig:comp_to_total}). 

\item \textit{Wavelength range cut and Voronoi binning.} For the full analysis, we used the wavelength range between 4800 and 5500~$\si{\angstrom}$. Within the MUSE spectral coverage, this range contains the relevant spectral features sensitive to the stellar population parameters of interest, such as H${\beta}$ and several Mg and Fe absorption lines indicated in Fig.~\ref{fig:spectra} \citep[e.g.,][]{Pinna2019b, Pinna2019a, Sattler2023, Sattler2025}. At the same time, this range is not as affected as the redder range by sky residuals and imperfections in the modeling of molecular bands of the stellar population models. 
Using this wavelength range, we applied a Voronoi binning \citep{Cappellari2003}, using the \texttt{VORBIN} Python package. This method spatially bins the data in an adaptive way, preserving the maximum possible spatial resolution, to reach a target signal-to-noise ratio (S/N) per bin. 
A mild level of spatial binning was initially required to reach the required S/N for an acceptable level of accuracy in fitting kinematics, needed for step 3 (see below). 
We used a target S/N~$=25$, a good compromise between obtaining reliable velocities and velocity dispersion and preserving a close-to-spaxel resolution, especially in the very central region where our interest resides. Preserving the highest possible spatial resolution is important for the spectro-photometric decomposition, and binning more would have smoothed the spatial information that C2D needs to fit each of the quasi-monochromatic images. 
We ensured that all included spaxels before binning had a S/N$>3$. 

\item \textit{Correction from stellar kinematics.} C2D performs the spectro-photometric decomposition in each individual quasi-monochromatic image, which is created with the flux in the same spectral pixel, i.e. at the same wavelength position, for all spaxels. 
When extracting the images, the spectral features in the data cube (e.g. absorption or emission lines) must be located in the same spectral pixels (same wavelength) for all spaxels. 
Otherwise, one image may mix information from different rest-frame wavelengths. 
In the observed spectra, spectral features are shifted not only an amount corresponding to the redshift of the galaxy, but also due to (gas- and stellar-) kinematic Doppler effects. Moreover, the observed width of these spectral features is affected by the stellar and gas velocity dispersion. 
Therefore, before applying C2D, we need all spaxels to have the same velocity ($V=0$~km~s$^{-1}$ at restframe) and the same line broadening (velocity dispersion). For that, we corrected the data cube from the stellar kinematics of each spaxel. We calculated the LOS velocity and velocity dispersion in each spaxel of our cropped and Voronoi binned data cube, as explained in Sect.~\ref{sub:spectr_fit}. 
Accounting for the kinematics in each spaxel, we shifted the entire data cube to the rest frame and we convolved it to the maximum measured velocity dispersion in the analyzed data cube (${\sim}155$~km~s$^{-1}$). 

\item \textit{Treatment of the emission lines.} Since M~74 is a star-forming galaxy, spaxels in the disk region have emission lines, often strong, in their spectra. We tested C2D on the kinematic corrected data cube both with and without previously subtracting emission lines. The emission cleaning for these tests was conducted using \textsc{pPXF} (\citealt{Cappellari2004}, see also Sect~\ref{sub:spectr_fit}). 
We finally decided not to feed C2D with the emission-subtracted data cube, but rather use the cube including the emission. This choice avoided introducing any additional bias when subtracting emission lines from spectra, and gave better C2D fits. Since emission characterizes specific regions, such as the disk-dominated region, and not others like the very central, NSC-dominated region, preserving these differences helped C2D to perform a better multicomponent fit. 
\citet{MendezAbreu2019} showed the capability of C2D to process emission lines properly. 
\end{enumerate}
\subsubsection{Application of C2D to M~74} \label{sub:c2d_m74}
\begin{figure*}
\centering
\resizebox{0.85\textwidth}{!}
{\includegraphics[scale=1.]{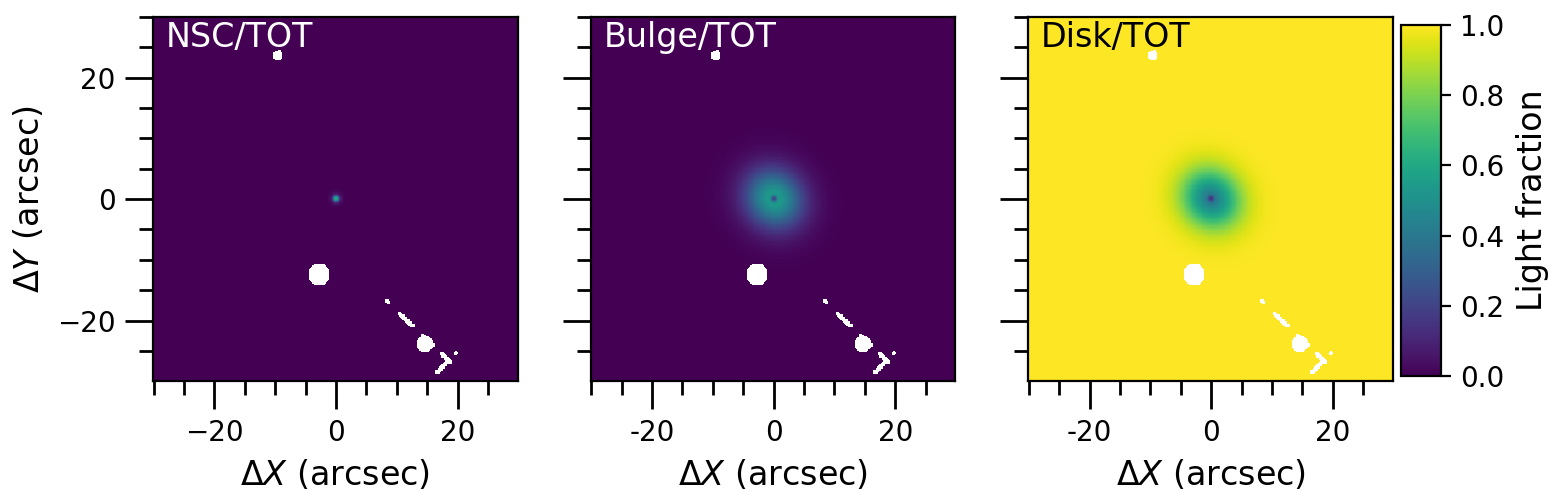}}
\caption{Light ratio of each component over the total light of the galaxy, delivered by C2D. From left to right: NSC fitted as a PSF, bulge and disk. The NSC dominates the light in the central spaxels within one PSF FWHM. Masked regions are plotted in white. 
}
\label{fig:comp_to_total}
\end{figure*}
This work analyzes the stellar populations of M~74’s NSC and compares them with those of the host galaxy. Because the NSC is unresolved, its light was measured within a PSF aperture, while the surrounding galaxy was spatially resolved using the full capabilities of the IFS data. This enables a direct comparison between the NSC and its immediate environment, placing its formation in the context of the host galaxy.
We adopt C2D over other spectrophotometric decomposition codes (e.g., BUDDI, \citealt{Johnston2017}, see also Sect.~\ref{sec:intro}), because it uniquely preserves both spectral and spatial information for each component. This allows us not only to subtract the PSF-weighted NSC light but also map the host-galaxy properties, fully exploiting the spatial information from IFS. 

For this work, we adapted C2D and GASP2D to include the following three components: an NSC, a bulge and a disk. This choice was, on the one hand, motivated by the fact that this galaxy was fitted with a bulge and a disk by \citet{Salo2015} as part of the S$^4$G sample \citep{Sheth2010}. On the other hand, we are mostly interested in the NSC, which is not spatially resolved in MUSE data as its size is smaller than the PSF size (Sect.~\ref{sub:galaxy_nsc}). Its light is spread by the PSF in a larger region than its size, and we follow other studies modeling the NSC as a PSF to disentangle its spectra from the galaxy spectra \citep[e.g.,][]{Johnston2020, Fahrion2021}. 
We fitted the NSC using the Moffat function from the PSF modeling in \citeauthor{Emsellem2022} (\citeyear{Emsellem2022}, see Sect.~\ref{sec:obs} for more details). While we adopted a FWHM of 0.6~arcsec (three MUSE spaxels), we tested that the results were not affected by variations of $\pm 0.2$~arcsec (one spaxel). 
To test this, we repeated the full decomposition and spectral-fitting process with a FWHM of 0.4 and 0.8~arcsec, finding that the differences in the NSC stellar populations were negligible. 
The FWHM and the power index (2.8, see Sect.~\ref{sec:obs}) of the Moffat function were fixed in the fitting process, while the PSF (NSC) central intensity was kept free and fitted. 

For the bulge and the disk, we used a Sérsic and a broken-exponential two-dimensional profiles, respectively (see \citealt{MendezAbreu2017} for the specific implementation). 
We extensively tested different combinations of free and fixed parameters of these profiles, resulting in a better decomposition with a certain number of them fixed, as described below. Furthermore, the impact of small errors in the bulge and disk parameters on the NSC spectra, which is the goal of the decomposition, is negligible. 
The effective intensity, the effective radius and the Sérsic parameter of the bulge, as well as the central intensity of the disk, were kept free and fitted by the code, while the ellipticity and the position angle of both the bulge and the disk were fixed to the values from the S$^4$G decomposition by \citet{Salo2015}, who fitted the disk with a single exponential. 
We checked that the effective radius and the Sérsic parameter of the bulge resulting from the fit were very similar to the values from \citet{Salo2015}. 
We did not expect to find the same values since they used a different wavelength range compared to ours. 

Before choosing a broken exponential for the disk profile, we performed a preliminary morphological analysis, fitting ellipses to the isophotes, in an image obtained from the entire PHANGS-MUSE mosaic, integrating all wavelengths in the analyzed spectral range. 
A one-dimensional radial profile, obtained by averaging azimuthally, follows a broken exponential, with the outer-profile scale length matching the disk scale length from S$^4$G \citep[using a single exponential]{Salo2015}. This is in agreement with the fact that the S$^4$G team performed the morphological fit on a much larger spatial scale. We measured that the inner exponential has a scale length of 17.4~arcsec (${\sim} 827$~pc) while the break radius is located at 30~arcsec (${\sim} 1.4$~kpc) from the center. While our cropped data cube is dominated by the inner exponential, we fixed both these parameters in the C2D decomposition. 
With our new setup, GASP2D fitted each quasi-monochromatic image with a combination of a bulge, a disk and a NSC. 
This resulted in three data cubes, one for each component. Images resulting from integrating all wavelengths for each data cube are shown in Fig.~\ref{fig:datacubes}.
The light fractions of each component over the total are shown in two dimensions in Fig~\ref{fig:comp_to_total}. This figure shows that the NSC dominates the light (with a light fraction above 60\%) in the central spaxels within one PSF FWHM, while 100\% of the light was assigned to the disk from a certain radius (${\sim}7$~arcsec) outwards. 

Since we are interested in the stellar-population properties of the NSC and its comparison with the surrounding host galaxy, we will use for the rest of the analysis a data cube for the NSC and another for the rest of the galaxy. The latter was obtained by combining the bulge and the disk data cubes, and corresponds to the host galaxy, i.e. the galaxy after subtracting the NSC. We preferred this approach instead of considering the bulge and the disk separately to minimize the impact of a potential suboptimal decomposition of these two morphological structures. 
This also follows previous studies of this kind, comparing the NSC to the host galaxy as a whole \citep{Johnston2020, Fahrion2021}. 
However, the C2D approach, providing spatially resolved spectral information, additionally allows us to fairly compare NSC properties with the host galaxy in the same aperture. This minimizes biases derived from quantitative comparisons of different spatial regions, due to spatial gradients in the host galaxy. 

\begin{figure}
\centering
\resizebox{0.47\textwidth}{!}
{\includegraphics[scale=1.]{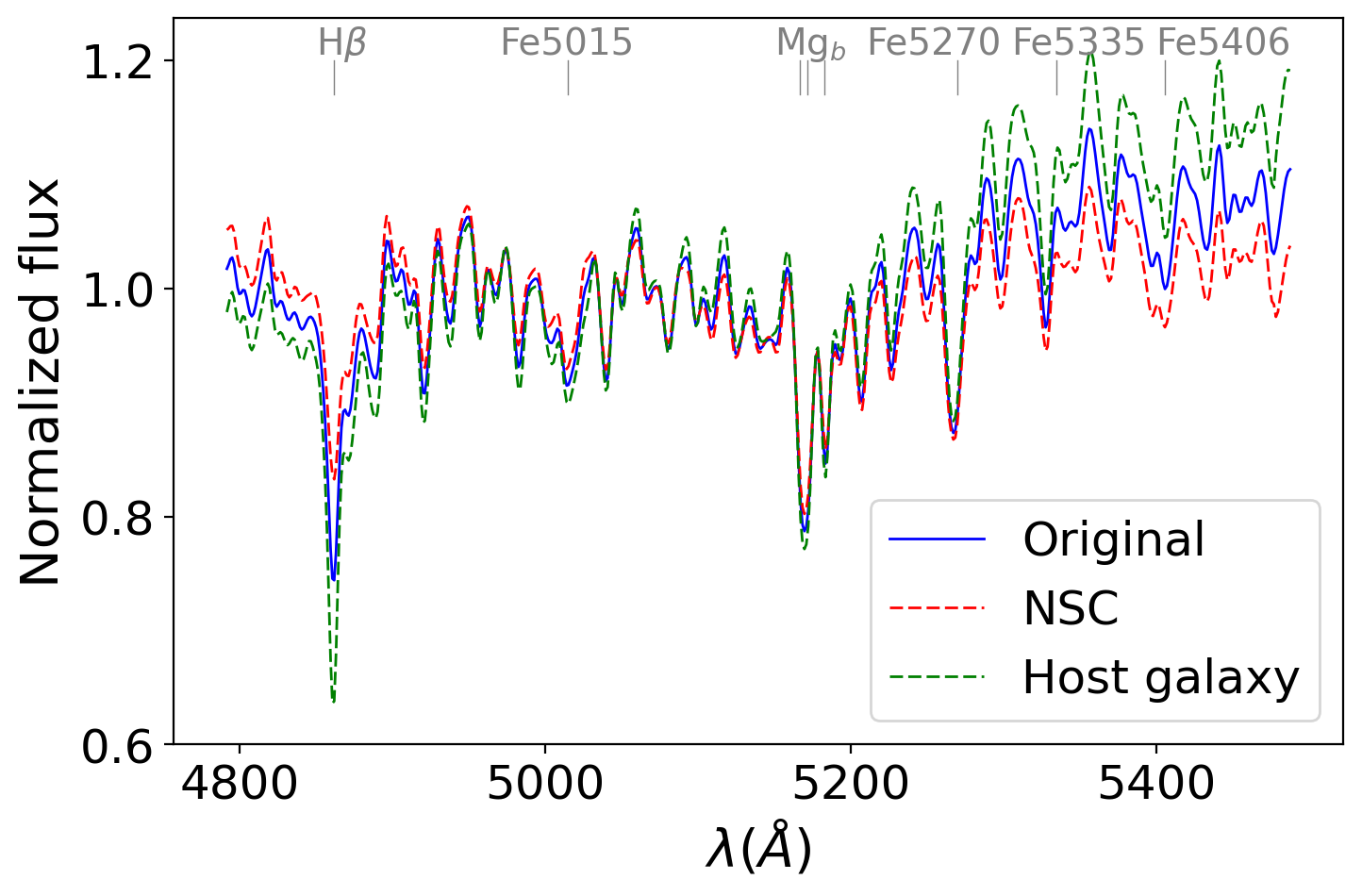}}
\caption{Spectra obtained by integrating, within an aperture of $1\sigma$ of the PSF, the prepared, kinematic-corrected ({\it original}) data cube before the spectrophotometric decomposition ({\it blue solid}), the {\it NSC} data cube ({\it red dashed}), and the {\it host-galaxy} data cube ({\it green dashed}). Some spectral features, relevant for the stellar population analysis, are shown: H$\beta$, the Mg$_b$ triplet (not entirely resolved), and four Fe lines.
}
\label{fig:spectra}
\end{figure}

We extracted integrated spectra in an aperture of radius $1\sigma$ of the PSF FWHM (${\sim}0.26$~arcsec $\approx 12$~pc) from these two data cubes and, for comparison, from the ''original'' kinematic-corrected data cube before being processed by C2D (Sect.~\ref{sub:preparation}). 
Taking into account the central region of the PSF ensures that small errors in its estimate and variations of its size with wavelength have no impact on the extracted properties of the NSC. 
Hereafter, we will refer to: (1) the NSC spectrum and data cube as the one corresponding to the NSC component as extracted by C2D, (2) the ''host-galaxy'' spectra and data cube as the one obtained combining the C2D disk and the bulge (and corresponding to the galaxy after subtracting the NSC), and (3) the ''original'' spectrum and data cube as the original one before the C2D decomposition and containing the three morphological components (i.e., the full galaxy). 
These spectra are shown and compared to each other, for $1\sigma$~PSF, in Fig.~\ref{fig:spectra}. 

\begin{table*}[!h]
\caption{\label{tab:SP} Average stellar-population properties and their uncertainties, 
obtained as detailed in Sect.~\ref{sub:spectr_fit} from spectra integrated in a $1\sigma$ PSF aperture. From top to bottom: the NSC, the host galaxy (decomposed as explained in Sect.~\ref{sub:decomp}), and all components combined in the LOS (obtained from the original - kinematic-corrected - data cube). From left to right, columns indicate: light-weighted (LW) average age, total metallicity [M/H] and [Mg/Fe] abundance; mass-weighted (MW) average age, [M/H] and [Mg/Fe]. 
Uncertainties were obtained via Monte Carlo simulations (Sect.~\ref{sub:spectr_fit}). 
}
\centering 
\begin{tabular}{|c|ccc|ccc|}
\hline\hline
 & Age$_{\rm{LW}}$ (Gyr)& [M/H]$_{\rm{LW}}$ (dex) & [Mg/Fe]$_{\rm{LW}}$ (dex)  & Age$_{\rm{MW}}$ (Gyr)& [M/H]$_{\rm{MW}}$ (dex) & [Mg/Fe]$_{\rm{MW}}$ (dex)\\
\hline
NSC     & $11.1\pm0.5$  & $-0.55 \pm 0.02$ & $0.07 \pm 0.03$ & $11.5 \pm 0.5$ & $-0.49 \pm 0.03$ & $0.03 \pm 0.02$ \\
Host  & $4.7 \pm 0.4$  & $0.24 \pm 0.02$ & $0.03 \pm 0.01$ & $9.1 \pm 0.7$  & $0.25 \pm 0.03$  & $0.15\pm 0.02$ \\
Original & $6.8 \pm 0.7$  & $-0.23 \pm 0.03$  & $0.07 \pm 0.02$ & $9.6 \pm 0.8$ & $-0.27 \pm 0.03$ & $0.02 \pm 0.03$\\
\hline\hline
\end{tabular}
\end{table*}
\subsection{Spectral fitting} \label{sub:spectr_fit}

We used a full-spectrum fitting technique both for the stellar population analysis (of all previously mentioned integrated spectra and data cubes), and for the calculation of the kinematic parameters necessary for their correction, as explained in Sect.~\ref{sub:preparation} (step 3). We used 
the latest Python implementation of the penalized pixel-fitting (\ppxf) method \citep{Cappellari2004, Cappellari2017, Cappellari2023}. \ppxf{} models an observed spectrum as a linear combination of simple stellar population (SSP) templates, convolved with a line-of-sight velocity distribution (LOSVD) described by a Gauss–Hermite series. The resulting (light- or mass-fraction) weights, assigned to the SSP models to obtain the best-fit spectrum, can be used for the stellar-population analysis, while the LOSVD can be used to extract the kinematic parameters. 
In this work, different \ppxf{} setups were used to extract stellar populations and kinematics. 
Stellar velocities and velocity dispersions were derived by fitting the central data cube resulting from step 2 in Sect.~\ref{sub:preparation}, with \ppxf{}, using an additive polynomial of degree 8, and no multiplicative polynomial. Additive polynomials are typically used to minimize the effect of template mismatch in the kinematic fitting \citep{Cappellari2004,Cappellari2017}. 

The stellar population analysis was carried out applying \ppxf{} to the NSC and host-galaxy data cubes and PSF-integrated spectra, resulting from the C2D decomposition (Sect.~\ref{sub:c2d_m74}). For comparison, the ''original'' data cube (before the C2D decomposition, but kinematic corrected) was also analyzed in the same way. The \ppxf{} setup used for the stellar populations included a multiplicative polynomial of 8th order. Multiplicative polynomials are typically used to minimize the effect of imperfections in the flux calibration and dust reddening \citep{Cappellari2004,Cappellari2017}. On the other hand, additive polynomials are typically avoided in the calculation of ages and metallicities as they could affect absorption line strengths and bias the results \citep[e.g.,][]{Guerou2016}.
While the emission lines are strong in most disk-dominated regions of this star-forming galaxy, there is no clear emission in the very central region (Fig.~\ref{fig:spectra}). The H$\beta$ absorption line is an essential age indicator in the MUSE wavelength range \citep[e.g.,][]{Worthey1994}, and we decided not to mask it in the analysis of the PSF size of the data cubes (results in Sect.~\ref{sub:res:psf}). 
However, we needed to mask H$\beta$ in the analysis of the host-galaxy data cube (results in Sect.~\ref{sub:res:host}), since in the disk-dominated region the emission is strong.  
In all cases, we masked the other typical emission lines in the analyzed wavelength range ([O\,\textsc{iii}]$\lambda$$\lambda$4959,5007 and [N\,\textsc{i}]$\lambda$$\lambda$5198,5200). 

We used in this work the semi-empirical sMILES SSP models \citep{Knowles2023}, the most up-to-date version of the MILES SSP models \citep{Vazdekis2010, Vazdekis2015}. sMILES models are based on BaSTI isochrones \citep{Pietrinferni2004, Pietrinferni2006} and offer five different values of the [Mg/Fe] abundances, covering the range between -0.2 and 0.6~dex (in steps of 0.2~dex). Throughout this paper, we will use [Mg/Fe] as a proxy for the enhancement in $\alpha$ elements. 
sMILES models cover ages between 0.03 and 14~Gyr, with 53 different values, and ten values of total metallicities [M/H] between -1.79 and 0.26~dex. This gives a total of 2650 SSP models. 
Since each of them corresponds to one combination of age, [M/H] and [Mg/Fe], coupling these models with \textsc{pPXF} allows us to extract the distribution of these three stellar-population parameters from each spectrum (given by the distribution of weights given to the different SSPs). 
These models cover a wavelength range between 3540
and 7409.6~\si{\angstrom}, at a spectral resolution of
2.51~\si{\angstrom} at FWHM. 
Among sMILES sets of models, we chose the set of SSPs that adopted a bimodal initial mass function with a logarithmic slope of 1.3 for the segment of massive stars \citep{Vazdekis2003}. 
These models were normalized to correspond to one solar mass each, so that mass-weighted results are obtained from \textsc{pPXF} if SSPs are not renormalized. They need to be renormalized to their own median to obtain light-weighted results. We used both normalizations in this work, to obtain light-weighted stellar population maps and average results, and mass-weighted results to allow us to reconstruct SFHs in terms of mass fractions corresponding to different stellar populations. 

Our general approach for the stellar-population analysis was to average results from Monte Carlo simulations of unregularized \ppxf{} fits, following \citet{Cappellari2023}. 
We used this approach to fit the three spectra extracted from the PSF aperture (Sect.~\ref{sub:c2d_m74}), i.e. the NSC, the host-galaxy, and for comparison the spectrum from the original integrated data cube. 
We performed 100 Monte Carlo realizations for each spectrum. For each realization, we perturbed the spectrum with random, wavelength-dependent noise drawn from a Gaussian distribution. The noise level was taken from the residuals at the specific wavelength, from an initial (unregularized) \ppxf{} fit. 
From the distribution of SSP weights for the 100 realizations, we estimated the mean ages, [M/H] and [Mg/Fe], and their statistical uncertainties (as standard deviations), for each of the three spectra. This allowed us not to use any regularization on the SSP weights during the \ppxf{} fits. 

The resulting host-galaxy data cube from the decomposition was rebinned to a higher S/N of 100 before performing the spectral fitting for the spatially resolved stellar population analysis, ensuring robust results and lower uncertainties in the outer, fainter region. 
This still left us with 6766 Voronoi bins, which made the Monte Carlo approach time-consuming and computationally expensive for the entire data cube. Thus, we decided to map the stellar-population parameters of the host galaxy using \ppxf{} regularized fits with a unit regularization parameter. 
While one single unregularized fit gives discrete and noisy solutions, some level of regularization ensures smoother and more realistic stellar-population solutions \citep[e.g.,][]{Cappellari2017}. However, regularization above a certain level can lead to biases in the results and dilute features in the SFHs. Thus, we used the approach presented in \citet{Pinna2019a} to choose the regularization parameter (described as follows, see also \citealt{Pinna2019b, Martig2021, Sattler2023, Sattler2025}). 
After rescaling the \ppxf{} noise parameter as the one necessary to obtain a reduced $\chi^2=1$ with no regularization, we choose a regularization well below a certain maximum calibration value. We defined the calibration regularization as the minimum value increasing $\chi^2$, from the case with no regularization, by $\Delta \chi^2 \approx  \sqrt{2N_{\text{goodpix}}}$, where $N_{\text{goodpix}}$ is the number of fitted (spectral) pixels \citep{McDermid2015, Cappellari2017, Boecker2020}. 
This approach was also used to map the stellar populations from the original data cube before the C2D decomposition (Appendix~{\ref{app:res:orig_integr}}).  


\section{Results} \label{sec:results}

\subsection{Stellar populations within a central PSF aperture: average properties and SFHs of the NSC compared with the host galaxy} \label{sub:res:psf}
We present here the average stellar population properties and SFHs, both reconstructed from the weights given to the SSP models by the spectral-fitting approach detailed in Sect.~\ref{sub:spectr_fit}. All these results were extracted from the three spectra integrated within an aperture of $1\sigma$~PSF (Sect.~\ref{sub:c2d_m74}) and shown in Fig.~\ref{fig:spectra}. 

We present in Table~\ref{tab:SP} the average light- and mass-weighted stellar population properties of the NSC and the host galaxy and their uncertainties, calculated as a weighted average using the light or mass weights from the spectral fitting. For reference, we also present the results from the original integrated data cube, before the C2D decomposition. 
While we show in Table~\ref{tab:SP} both light- and mass-weighted results, differences between different components are enhanced in the light-weighted results, which in general give more weight to younger stars. 
The NSC is very old, with an average light-weighted age of ${\sim}11$~Gyr. It is metal poor ([M/H]$_{\rm{LW}}{\sim}-0.6$~dex) and slightly enhanced in $\alpha$ elements with respect to the Sun ([Mg/Fe]$_{\rm{LW}}{\sim}0.1$~dex). 
The host galaxy, in the central $1\sigma$~PSF aperture, shows much younger (light-weighted) ages (${\sim}5$~Gyr), higher and supersolar metallicities ([M/H]$_{\rm{LW}}{\sim}0.2$~dex) and about solar (light-weighted) [Mg/Fe] abundances. 
These results reveal that the NSC is much older and more metal-poor than its host galaxy, suggesting that it was formed a long time ago, when the interstellar medium was not chemically enriched yet, and was not involved in the later active star formation of the host galaxy. 
[Mg/Fe] is only slightly more enhanced in the NSC than in the center of the host galaxy. Since the $\alpha$ enhancement is usually interpreted as a formation in a faster time scale \citep[e.g.,][]{Matteucci1994}, this suggests that the formation time scale of the NSC may have been slightly faster than the central region of the host galaxy. This is in agreement with NSC's old age and low metallicity, suggesting that NSC formation was relatively short in time. 

On the other hand, the difference between the mass-weighted ages of the NSC and its host galaxy is smaller (than when considering light-weighted ages), with the stars in the host galaxy being only about 2~Gyr younger than the NSC. The smaller age difference between the host galaxy and the NSC, in the mass-weighted averages, stems from an older host-galaxy age. This suggests that, while the disk light is dominated by young stars, it also contains a significant (mass) fraction of old populations. However, this is still a significant difference for mass-weighted results, which are expected to be generally biased towards older ages \citep[e.g., ][]{SanchezBlazquez2014a, Sattler2023, Sattler2025}. Mass-weighted and light-weighted metallicities are quite similar to each other, while light-weighted [Mg/Fe] abundances are slightly higher than the mass-weighted ones. 
Uncertainties indicated in Table~\ref{tab:SP} were obtained via Monte Carlo simulations (Sect.~\ref{sub:spectr_fit}). Differences in the stellar-population properties of the different components are larger than these uncertainties and thus they can be considered significant. However, these represent only statistical uncertainties and they are relatively low for such high S/N spectra. 
Systematic uncertainties, such as those due to imperfections in the SSP models, are very challenging to estimate and are not taken into account here. 

\begin{figure}[!]
\centering
\resizebox{0.45\textwidth}{!}
{\includegraphics[scale=1.]{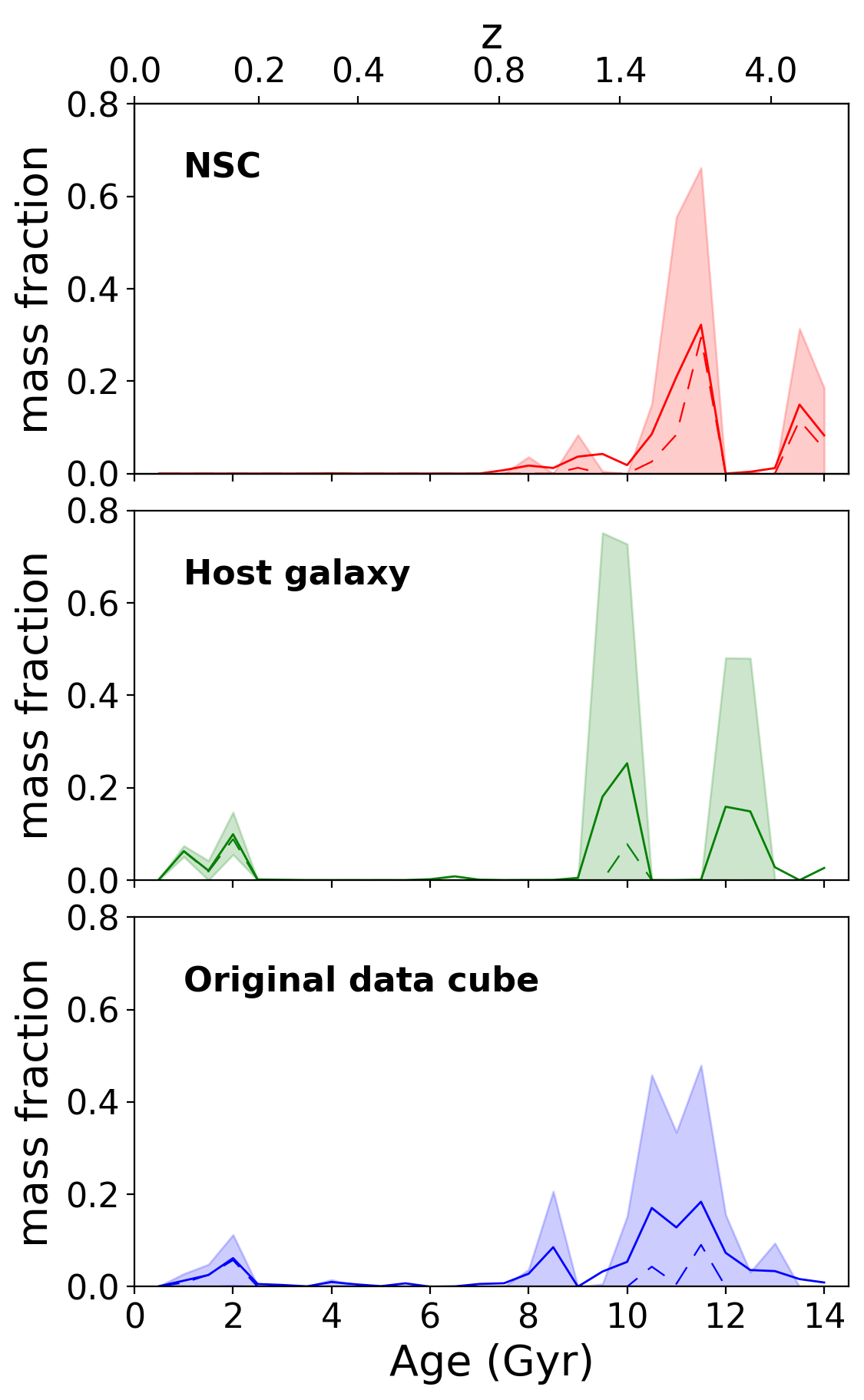}}
\caption{SFHs of the {\it NSC}, the center of the {\it host galaxy}, and both components together (from the {\it original} MUSE data cube), from top to bottom. The three panels were plotted using the mass-weighted results from fitting spectra integrated in a central PSF aperture (Sect.~\ref{sub:c2d_m74} and Fig.~\ref{fig:spectra}). Mass fractions are normalized to the mass of each component within the PSF aperture. 
Solid lines indicate the mean mass fraction for each age bin (half Gyr wide), calculated on the weight distributions from the 100 Monte Carlo realizations (Sect.~\ref{sub:spectr_fit}). Dashed lines follow the median, and shades indicate the 16\% and 84\% percentiles from the distributions. 
}
\label{fig:sfh}
\end{figure}
\begin{figure}[!]
\centering
\resizebox{0.5\textwidth}{!}
{\includegraphics[scale=1.]{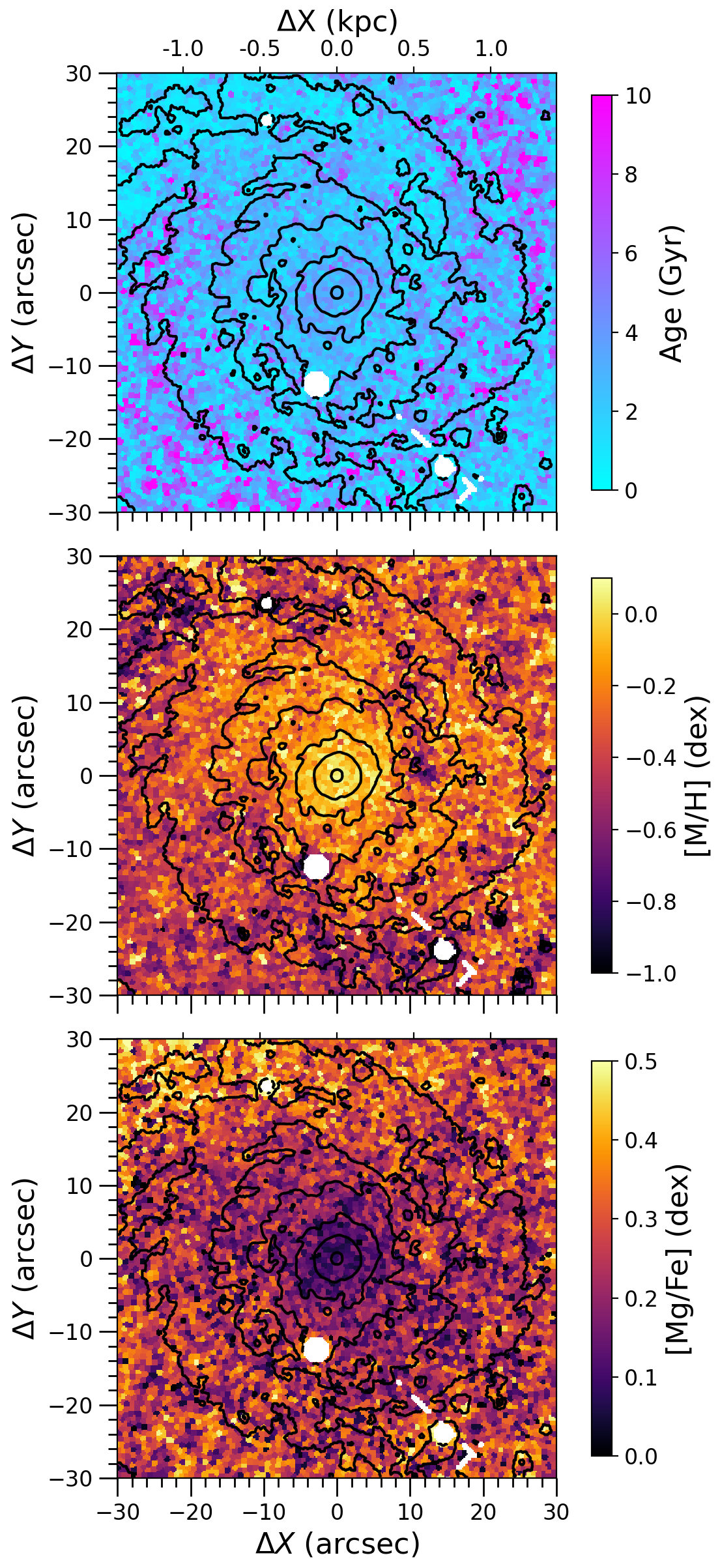}}
\caption{Stellar population maps of the host-galaxy component of M~74, after subtracting the NSC. From top to bottom: light-weighted mean age, total metallicity [M/H] and [Mg/Fe] abundance. Masked regions are depicted in white and isophotal contours in black. The physical scale is given as a reference in the top $X$ axis. 
}
\label{fig:host_pop_maps}
\end{figure}

From the mass weights obtained in the full spectral fitting process, we also reconstructed the SFHs in terms of the stellar mass assigned to different age bins. Compared to the average properties presented above, SFHs were obtained by averaging only the mass weights given to coeval SSPs. 
We show these SFHs in Fig.~\ref{fig:sfh}, respectively in the top and middle panels. In the bottom panel, we show, as a reference, the SFH of the full non-decomposed galaxy, extracted from the same PSF aperture in the kinematic-corrected original MUSE data cube before the C2D decomposition (blue spectrum in Fig~\ref{fig:spectra}). 
The NSC shows distinct stellar populations at old ages, with the highest mass fractions distributed in ages between 10 and 11~Gyr. 
The host galaxy shows an old component made of two peaks of different ages (${\sim}10$ and ${\sim}12$~Gyr) and a relatively recent component with ages of about 2~Gyr and younger. 
Some intermediate-age populations are also present (${\sim}6$~Gyr). 
The SFH of the full galaxy (bottom panel) shows a combination of all these components. 
A word of caution is deserved by the large uncertainties, indicated by the shades in each panel of Fig.~\ref{fig:sfh}. These shades cover the region between the 16\% and 84\% percentiles, calculated on the distribution of the 100 Monte Carlo realizations. Since mass fractions are normalized to the total mass in each SFH (in each fitted spectrum), these errors indicate that some mass fraction assigned to the highest peaks could be, in reality, redistributed to minor peaks. 
Overall, the SFHs suggest that the center of the galaxy had an intense early phase of star formation, between 14 and 8~Gyr ago, when the NSC stars formed. 
Afterwards, it experienced a long quiescent phase, with a very low star-formation rate. Only relatively recently (later than 3~Gyr ago), a more active star formation was reactivated in M~74. However, while this episode involved an extended region of the disk and was enhanced in spiral arms, as shown in Fig.~\ref{fig:host_pop_maps}, it did not affect the NSC, as this remained old and metal poor. 
\citet{SanchezBlazquez2014b} found a very similar SFH for M~74, using the STEllar Content via Maximum A Posteriori likelihood
(STECKMAP) spectral-fitting code \citep{Ocvirk2006}. 

\subsubsection{Additional tests and comparisons}\label{sub:tests}
We first compare the results above with the stellar populations obtained from a central PSF aperture in the kinematic-corrected original data cube (spectrum indicated as a blue solid line in Fig.~\ref{fig:spectra}). 
This contains the integrated light of the NSC and the host galaxy together and, as expected, gives intermediate stellar populations between the NSC and the host galaxy (Table~\ref{tab:SP}). 
We can also compare the stellar populations in a central PSF aperture with the surrounding region of the galaxy, 
both extracted from the original data cube. For that, we calculated the average ages, [M/H] and [Mg/Fe] within a much larger central aperture of 100~pc (excluding the central $3\sigma$~PSF to make sure the contamination from the NSC is negligible). For this, we used the stellar-population maps extracted from the original integrated data cube (Appendix~\ref{app:res:orig_integr}). 
These stars are younger (5.6~Gyr) and slightly more metal-rich ($-0.1$~dex), than the stars in the central PSF aperture extracted from the same original data cube (Table~\ref{tab:SP}). 
However, the [Mg/Fe] abundance is very similar (0.1~dex). 
The fact that the original data cube, including the light of stars of both the NSC and the host galaxy, shows gradients with older and more metal-poor stars in the very center, is in perfect agreement with the presence of an older and more metal-poor component in the center (i.e. the NSC), which is not present in the surroundings. 

We have also tested the robustness of this gradient by fitting a spectrum from the original data cube, but integrated now in a $3\sigma$~PSF (about 40~pc, with a much stronger contribution from the host galaxy). 
In a $3\sigma$~PSF aperture, we find ages about 1~Gyr younger than integrating within $3\sigma$~PSF, in the original and host-galaxy data cubes, while the NSC age remains the same. 
Since we are comparing results from slightly different methods (on the one hand, fitting spectra integrated in specific apertures, on the other hand, averaging stellar-population properties in the spatially resolved maps shown in Appendix~\ref{app:res:orig_integr}), we have checked that the same results were found within $1\sigma$~PSF with the different methods. The mean age, [M/H] and [Mg/Fe] from a $1\sigma$~PSF aperture in the spatially resolved maps in Fig.~\ref{fig:orig_pop_maps} are respectively: $6.8$~Gyr, $-0.2$ and $0.1$~dex, approximately the same as the results from fitting a spectrum from the original data cube obtained integrating in a $1\sigma$~PSF aperture (shown in Table~\ref{tab:SP}). 

To additionally test how robust our C2D method is, we have also compared our results with the ones from an NSC spectrum obtained by subtracting the underlying host-galaxy light from a PSF integrated spectrum (''original'', in Fig.~\ref{fig:spectra}). Following \citet{Fahrion2021}, the host-galaxy spectrum was obtained from an annulus surrounding the NSC, with inner and outer radii of 0.8 ($\sim 3\sigma$~PSF) and 2 arcsec. 
Before subtracting the host-galaxy spectrum from the ''original'' spectrum, the annulus spectrum needs to be first rescaled to the light of the host galaxy in the same central PSF aperture where the ''original'' spectrum was integrated (brighter than the annulus). While \citet{Fahrion2021} performed a dedicated morphological decomposition to calculate the scaling factor, we used our decomposition from C2D. We obtained again very old light-weighted ages ($8.5 \pm 1.1$~Gyr), low metallicities ($-0.39 \pm 0.04$~dex) and close to solar [Mg/Fe] enhancement ($0.01 \pm 0.03$~dex). While these results give small differences (taking into account the uncertainties - slightly younger ages and higher metallicities), they are closer to the results from C2D than what was obtained from the original data cube. While it is not surprising that different methods give slightly different results (see also Sect.~\ref{sub:disc:nsc}), our method is more sophisticated than simply subtracting spectra. Our method leads to a cleaner NSC spectrum than the simple subtraction, and therefore to larger differences between results from integrated and decomposed spectra. The main reason for this is that C2D performs a fitting spaxel by spaxel and spectral pixel by spectral pixel, while when subtracting the annulus spectrum from the PSF spectrum we are forced to take an ''average'' spectrum within the annulus, and the scaling factor is also calculated using integrated light. 
More importantly, these differences are not critical for the interpretation of our results: the NSC appears to be very old and metal poor, and much older and more metal poor than the host galaxy according to the results obtained with all methods.

\subsection{Spatially resolved stellar populations of the host galaxy} \label{sub:res:host}

Spatially resolved properties of the host galaxy are important to interpret the NSC properties in the context of its surroundings. 
We mapped the mean age, [M/H] and [Mg/Fe] of the host galaxy, fitting the host-galaxy data cube decomposed with C2D (Sect.~\ref{sub:c2d_m74} and ~\ref{sub:spectr_fit}). 
The maps are shown in Fig.~\ref{fig:host_pop_maps}. 
Both [M/H] and [Mg/Fe] are presented in logarithmic scale and with respect to solar values (zero values correspond to solar values). 
The age map (top panel) shows a relatively young host galaxy (with most regions younger than ${\sim}4$~Gyr). A pattern following the spiral arms can be identified, with alternative regions dominated either by young (about ${\sim}2$~Gyr-old or younger) or old stars (older than ${\sim}8$~Gyr). 
The [M/H] map (middle panel) displays a clear negative radial gradient, with a central metal-rich region of about 500~pc size (supersolar metallicities). Outer regions have higher and lower values alternatively.  
[Mg/Fe] abundances are mostly anti-correlated with [M/H], showing the lowest values (close to solar) in the central metal-rich ${\sim}500$~pc. 
Hints of the spiral-arm shape are visible in both [M/H] and [Mg/Fe] maps, whose values do not necessarily follow any correlation or anti-correlation with age. 
However, the youngest (outer) regions in the analyzed data cube show a strong $\alpha$ enhancement in some metal-poor blobs.

Masking H$\beta$ might introduce an age bias in the map of the host galaxy, with respect to the analysis of the central PSF size, in which we directly fitted H$\beta$ since there was no visible emission (see Sect.~\ref{sub:spectr_fit} for details). 
However, we verified that fits of the host-galaxy PSF-sized spectrum give similar results, no matter whether H$\beta$ is masked or not. 
Thus, we can qualitatively compare the results in Sect.~\ref{sub:res:psf} with the maps shown here. 
The NSC is as old as some outer interarm disk-dominated regions in the host-galaxy age map in Fig.~\ref{fig:host_pop_maps}, while the rest of the host galaxy is much younger. 
The old metal-poor NSC resides in the relatively young and very metal-rich central ${\sim}500$~pc of the host galaxy. 

However, the NSC is as metal-poor as some young regions of the disk. 
Regarding [Mg/Fe], the NSC has a similar abundance to its surrounding galaxy. 
These results suggest that while M~74 kept forming stars in most of its disk, this did not happen in some specific regions. While more recent star formation occurred in the central region of the galaxy, allowing for an extended chemical enrichment, this did not happen in the NSC, which remained old and metal-poor. 

In Appendix~\ref{app:res:orig_integr}, we present for reference the analysis of the original MUSE data cube before the spectro-photometric decomposition. Although the Voronoi binning is slightly different, Fig.~\ref{fig:orig_pop_maps} shows similar trends to Fig.~\ref{fig:host_pop_maps}. The difference between the original data cube and the host-galaxy decomposed data cube is that the former includes the light of the NSC, which was extracted from the latter. Thus, we can expect a difference in the stellar populations only in the region where the NSC dominates. Indeed, as described in Appendix~\ref{app:res:orig_integr}, a clear [M/H] drop is visible within a PSF size in Fig.~\ref{fig:orig_pop_maps}. 
Our results for the central region of M~74 are in agreement with the stellar-population analysis by \citeauthor{Pessa2023} (\citeyear{Pessa2023}, see also Sect.~\ref{sec:galaxy}).

\section{Discussion} \label{sec:disc}

\subsection{The nuclear star cluster in M~74 and its formation} \label{sub:disc:nsc}

We have found old ages (${\sim}11$~Gyr), low metallicities ([M/H]${\sim}-0.55$) and no particular $\alpha$ enhancement ([Mg/Fe]${\sim}0.1$) in the NSC hosted by the relatively massive spiral galaxy M~74. 
We found older ages and lower metallicities in the NSC compared to results from \citeauthor{Hoyer2023b} (\citeyear{Hoyer2023b}, $8 \pm 3$~Gyr and $Z=0.012 \pm 0.006$, corresponding to [M/H]${\sim}-0.2$~dex). However, our results for the decomposed NSC are compatible within their large uncertainties. They used a different method and wavelength range (SED fitting of HST and JWST photometry, see also Sect.~\ref{sub:galaxy_nsc}). They did not decompose the light of the NSC from the host galaxy in the LOS, but they could spatially resolve the NSC. These reasons can explain the differences, as they found ages and metallicity values more similar to our results before the spectro-photometric decomposition (from the ''Original'' spectrum, in Table~\ref{tab:SP}). 

Our stellar population analysis has shown an ancient and metal-poor NSC, surrounded by a much younger and more metal-rich region of the host galaxy. As shown by JWST observations \citep{Hoyer2023b}, the NSC lives in a cavity with no gas or dust, while being hosted by an actively star-forming galaxy with a complex structure of gas and dust filaments. 
The cavity formed due to some processes either keeping gas and dust out of it or wiping them out of the central region (see also Sect.~\ref{sub:NSCform_host}). 
While the presence of a massive black hole in M~74 is still under debate, the presence of a low-luminosity AGN was proposed by different authors \citep[e.g.,][]{Dong2006, She2017}. It is also one of the possible explanations for the presence of the cavity (all gas may have been blown via AGN feedback, especially if the black-hole activity was much stronger in the past) and of the mid-infrared emission found at the center of the galaxy with JWST observations \citep{Hoyer2023b}. 
Similarly, central quiescent regions (of about 100~pc sizes), surrounded by starbursts, as well as nuclear older ages (within radii from few to ${\sim}50$~pc) than the circumnuclear regions, were previously detected in AGN hosts \citep[e.g.,][]{Marquez2025, daSilva2025}. 
Other, more unlikely explanations for the fact that no gas in an amount sufficient to trigger star formation reached the NSC in the last $\sim$8~Gyr, could be a previous overall scarcity of gas in the galaxy (not happening now since the galaxy is overall gas rich), and the lack of a clear and strong bar structure, which would help gas inflow, combined with dynamical processes making gas inflow inefficient. 

In any case, the presence of the cavity may have affected the evolution of the NSC, which could not grow via in-situ star formation due to the lack of gas. 
However, the surrounding region of the NSC, including stars in the region of the cavity, is much younger and more metal-rich. Our decomposition, supported by the analysis of spectra in the original data cube integrated into different apertures (Sect.~\ref{sub:tests}), suggests that these younger stars belong to the galaxy disk. Our SFH of the center of the host galaxy suggests that the latter had later and relatively recent episodes of star formation. This did not happen in the NSC, whose stars were all formed at the early stages of the evolution of M~74. Its SFH shows that its stars were all formed more than 8~Gyr ago. Regarding the origin of these old and metal-poor stars, we cannot say if they were formed in situ a long time ago, when M~74's gas was still metal-poor. Old ages and low metallicities (although typically lower than what we find here) in NSCs are traditionally explained by globular cluster inspiral \citep{Neumayer2020, Fahrion2021, Fahrion2022a, Fahrion2022b}. While globular cluster migration is thought to be the dominant formation mechanism for NSCs less massive than the one in M~74 and hosted by low-mass galaxies, our results do not provide enough information to rule out (or support) this scenario versus a very early in-situ formation (see also Sect.~\ref{sub:NSCform_host}). 
Furthermore, the presence of LMXBs in the NSC suggested by the nuclear X-ray emission (see Sect.~\ref{sub:galaxy_nsc}), consistent with the NSC old age, provides additional support for the NSC to be one or more globular clusters that migrated to the galactic center.

\subsection{Is M~74's nuclear star cluster unique?} \label{sub:disc_unique}

The NSC in M~74 shows common stellar population properties with other NSCs in the literature. However, it is important to discuss NSC properties in the context of their host galaxies. 
\citet{Fahrion2021} analyzed NSCs in a sample of 25 ETGs in the Fornax and Virgo clusters, finding several of them as old as M~74's NSC and very metal-poor. 
However, while the sample includes galaxies of similar stellar masses to M~74 (Sect.~\ref{sec:galaxy}), the NSCs with the most similar stellar populations to M~74's NSC were lower-mass NSCs ($\lesssim 10^{6}$\msun) hosted by less massive, mostly quiescent ETGs ($\lesssim 10^{9}$\msun). ETGs of similar masses as M~74 host metal-rich NSCs. 
Among the seven dwarf ETGs in the Fornax cluster, analyzed by \citet{Johnston2020}, NSCs with similar ages to the one in M~74 are much more metal-poor. These NSCs and their dwarf host galaxies are systematically more metal-poor than in the M~74 case. 
M~74's NSC age is also similar to that of three NSCs from \citet{Fahrion2022b}, who analyzed a sample of nine star-forming dwarf LTGs. While these galaxies are star-forming and have morphological types more similar to M~74, they have much lower stellar masses (${\sim}10^{7}$ to ${\sim}10^{9}$\msun). These host galaxies and their NSCs also show systematically lower metallicities than M~74 and its NSC.
Some NSCs hosted by galaxies more similar to M~74 in terms of morphological types and masses were analyzed by \citet{Kacharov2018}. However, while some have similar metallicities to M~74's NSC (see also \citealt{Walcher2006}), they are all relatively young. 
While there are other spiral galaxies with NSC ages similar to M~74's, these galaxies are slightly earlier types than M~74 \citep{Rossa2006}. 

While previous studies in the literature show other NSCs with similar stellar-population properties, suggesting that M~74's NSC is not unique, no NSCs hosted by a relatively young, metal-rich, massive star-forming (late-type) spiral galaxy, have been reported to have such old ages. 
This suggests that the properties of M~74's NSC are not common in its host-galaxy context, and this NSC may have had a peculiar formation and evolution history. 
It is quite extraordinary that in such a gas-rich spiral galaxy, no significant amount of gas triggered any in-situ star formation in the NSC over the last $\sim$8~Gyr. The inefficiency of gas inflow to the nucleus in this gas-rich star-forming galaxy is what makes M~74 unusual, resulting in its ancient NSC.

\begin{figure}
\centering
\resizebox{0.5\textwidth}{!}
{\includegraphics[scale=1.]{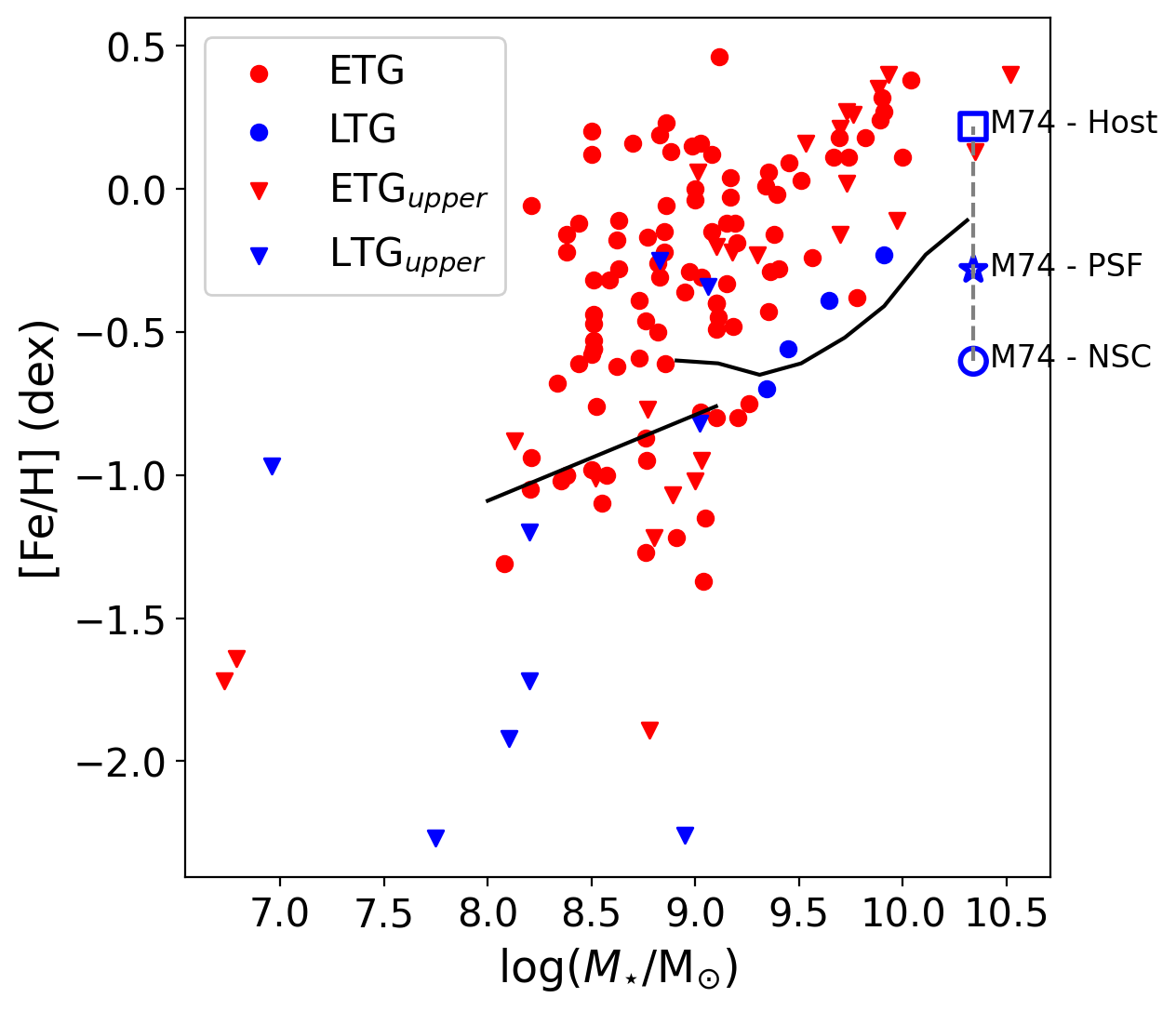}}
\caption{NSC metallicity as a function of the galaxy stellar mass (figure reconstructed and updated from Fig.~9 in \citet{Neumayer2020}). The blue open circle indicates the [Fe/H] metallicity of the decomposed M~74's NSC. For comparison, [Fe/H] of the decomposed host galaxy is indicated with an open square, while a star indicates the [Fe/H] from the original (integrated) data cube of M~74 (see the text for all the details). 
ETGs are indicated in red and LTGs in blue. Red circles are points from spectroscopic and photometric data \citep{Koleva2009, Paudel2011, Spengler2017, Kacharov2018}, as from \citet{Neumayer2020}. The two black solid lines are the mass-metallicity relations of galaxies, given as reference, at the low-mass end \citep{Kirby2013} and the high-mass end \citep{Gallazzi2005}. We have added the red triangles, which are NSC total metallicities ([M/H]) from \citet{Fahrion2021}, and are therefore considered as upper limits for [Fe/H]. We have also added all the blue points. Blue triangles are [M/H], and thus upper limits, of NSCs in dwarf LTGs in \citet{Fahrion2022b}. Blue circles are NSC [Fe/H] in more massive LTGs in \citet{Kacharov2018}. 
}
\label{fig:met_galmass}
\end{figure}

\subsection{NSC formation and the host-galaxy evolution}\label{sub:NSCform_host}

Different NSC properties have been found in galaxies of different masses (Sect.~\ref{sec:intro}). 
This was explained with a transition in the dominant mechanisms driving NSC formation and growth, with in-situ star formation being dominant in higher mass galaxies, and globular cluster migration to the center of a galaxy dominating at lower masses \citep[e.g., ][]{Fahrion2021, Fahrion2022a, Fahrion2022b}. 
Among others, one piece of evidence is the transition in the NSC metallicity regime at galaxy stellar masses of about $10^9$\msun\,\citep[e.g,][]{Neumayer2020}. In more massive galaxies, NSCs tend to be more metal-rich, with [Fe/H] metallicities about $\pm 0.5$~dex around solar. For lower galaxy masses, on the other hand, NSCs have metallicities in a wide range, including very low values. 
In Fig.{\ref{fig:met_galmass}}, we have reconstructed this trend as shown in Fig.~9 of \citealt{Neumayer2020}, and updated it with additional measurements (NSCs from \citealt{Fahrion2021, Fahrion2022b, Kacharov2018}). NSCs in LTGs are shown in blue, and show, on average, lower metallicities than NSCs in ETGs (red symbols). Note also that blue triangles are total metallicities ([M/H], from \citealt{Fahrion2022b}), and they have to be considered upper limits for [Fe/H] metallicities. While more points will be needed to shed more light on this trend, the available points suggest that the trend of more metal-rich NSCs being hosted by more massive galaxies, clearly shown for ETGs, also holds for LTGs (but with a lower average metallicity). For higher-mass spirals, we only have four NSCs from the literature (\citealt{Kacharov2018}, blue circles), which appear to display lower metallicities than, on average, NSCs in ETGs of similar masses. 

For M~74, [Fe/H] was calculated from [M/H] and [Mg/Fe] in Table~\ref{tab:SP} following the relation ${\rm [Fe/H]}={\rm [M/H]}-0.75{\rm [Mg/Fe]}$ from \citet{Vazdekis2015}. 
M~74's NSC (blue open circle), is more metal poor than the rest of NSCs in Fig.~\ref{fig:met_galmass} for similar galaxy masses, even if compared with the few LTGs. 
M~74 is located, in Fig.~\ref{fig:met_galmass}, in a mass regime where NSCs are thought to be dominated by in-situ formation, as suggested by their usually higher metallicities than their host galaxies \citep{Fahrion2022a}. 
However, its host galaxy (blue open square) is much more metal-rich than the NSC (about 0.7~dex difference in [M/H] within the central PSF region). 
In general, NSC old ages and low metallicities, in particular if older and more metal-poor than the host galaxy, have been interpreted as a main formation through globular cluster inspiral \citep[e.g.,][]{Fahrion2021}. 
If compared with other star clusters in M~74, its NSC shows similar ages to the most massive globular clusters \citep{Whitmore2023}, supporting the cluster migration scenario. 
On the other hand, models predict that NSCs of the mass of M~74's NSC (${\sim}10^7$\msun) assembled slightly more than half of their mass via in-situ star formation and the rest through globular cluster inspiral \citep{Fahrion2022a}. 
Recent JWST studies at high redshift have recently proposed that similar very dense objects existed already in the early universe and may have been the first structure that a galaxy formed \citep[e.g.,][]{Bellovary2025, Graham2025}. These high-redshift objects may have much in common with NSCs, ultra-compact dwarfs and stripped nuclei in the local universe. 
The old, metal-poor NSC in M~74, with very different stellar populations from the host galaxy, may have formed either by the later migration and merging of star clusters whose stars formed at very early stages, or by very early in-situ formation. 
Either a combination of both, several cluster accretion episodes, or different bursts of in-situ star formation might have led to the different peaks in the SFH in Fig.~\ref{fig:sfh}.

\section{Conclusions} \label{sec:concl}

We have analyzed the stellar populations of the NSC in M~74, and compared it with the host galaxy. 
We have used a spectro-photometric decomposition method, C2D, to separate the NSC light from the host galaxy. 
This method applies a morphological decomposition for different wavelengths, preserving variations both in the spatial and spectral directions. 
We performed a separate analysis, in a PSF aperture, of the NSC and the host galaxy, allowing us to compare them. 
Additionally, we have mapped the spatially resolved stellar-population properties of the central ${\sim}3$~kpc of the host galaxy. 
We have reported light- and mass-weighted ages, metallicities and [Mg/Fe] abundances, and reconstructed SFHs. 

We found an NSC average age of about 11~Gyr and a total metallicity [M/H] of about $-0.5$~dex. The SFH of the NSC shows that it is made up exclusively of old stellar populations. 
These results indicate that NSC stars formed during the earliest stages of the evolution of M~74, probably spanning 2-3 distinct episodes. While we cannot track the potential migration of these stars from other regions of the galaxy, and we did not find hints of their accretion from galaxy satellites, our results do not show later contributions with stars younger than ${\sim}8$~Gyr. The NSC appears to have remained inactive, in terms of star formation and chemical evolution, for a long time. 
Interestingly, this NSC lives in a cavity (with a size of ${\sim}300$~pc) devoid of gas and dust, while the rest of the galaxy shows a complex web of gas and dust filaments, where star formation is currently happening. 
Stellar-population maps of the host galaxy have shown young ages in the spiral arms, and a central very metal-rich ${\sim}500$~pc region, with lower [Mg/Fe] close to solar values. This indicates that the surroundings of the NSC kept evolving over time, while its evolution stopped.
The comparison of spectra extracted in a central PSF aperture, for the NSC and the host galaxy, shows that in contrast to the short NSC evolution, the center of the host galaxy had relatively recent star-formation episodes, in addition to the old populations coeval with the NSC. 

Finally, the NSC in M~74 is much more metal-poor not only than its host galaxy, but also than other NSCs hosted by ETGs of the same mass as M~74. While there are no measurements for spiral galaxies of the same mass as M~74, its NSC is more metal-poor than other NSCs in slightly lower-mass LTGs, with stellar masses between $10^{9.5}$ and $10^{10}$\msun. 
The literature measurements mentioned above might be contaminated by the light of the host galaxy. Our spectro-photometric decomposition method allowed us to obtain a cleaner view of the stellar populations of the NSC, otherwise contaminated by the host galaxy (as shown in Table~\ref{tab:SP} by the analysis of the ''Original'' data cube). 
While these findings would be traditionally interpreted as a dominant NSC formation through globular cluster migration and merger at the center of the galaxy, they also suggest that low metallicities might emerge from in-situ star formation (at early times) in massive LTGs. 
With the analysis of a larger sample of massive spiral galaxies, we will find out whether the NSC in M~74 is a unique case or if its stellar populations are common in star-forming galaxies of similar masses, providing additional insights on NSC formation.

\begin{acknowledgements}
FP acknowledges support from the Horizon Europe research and innovation programme under the Maria Skłodowska-Curie grant “TraNSLate” No 101108180, and from the Agencia Estatal de Investigación del Ministerio de Ciencia e Innovación (MCIN/AEI/10.13039/501100011033) under grant (PID2021-128131NB-I00) and the European Regional Development Fund (ERDF) ``A way of making Europe''. 
FP wishes to acknowledge the contribution of the IAC High-Performance
Computing support team and hardware facilities to the results of this research. 
MB acknowledges support by the ANID BASAL project FB210003. This work was supported by the French government through the France 2030 investment plan managed by the National Research Agency (ANR), as part of the Initiative of Excellence of Université Côte d’Azur under reference No. ANR-15-IDEX-01. This research was funded, in whole or in part, by the French National Research Agency (ANR), grant ANR-24-CE92-0044 (project STARCLUSTERS).  We thank the German Science Foundation DFG for financial support in the project STARCLUSTERS (funding ID KL 1358/22-1 and SCHI 536/13-1).
MQ acknowledges support from the Spanish grant PID2022-138560NB-I00, funded by MCIN/AEI/10.13039/501100011033/FEDER, EU. 
KG is supported by the Australian Research Council through the Discovery Early Career Researcher Award (DECRA) Fellowship (project number DE220100766) funded by the Australian Government. 
\end{acknowledgements}
\bibliography{fp_nsc_new}

\begin{appendix}
\appendix

\section{Analysis of the original MUSE data cube of the galaxy} \label{app:res:orig_integr}

\begin{figure}
\centering
\resizebox{0.5\textwidth}{!}
{\includegraphics[scale=1.]{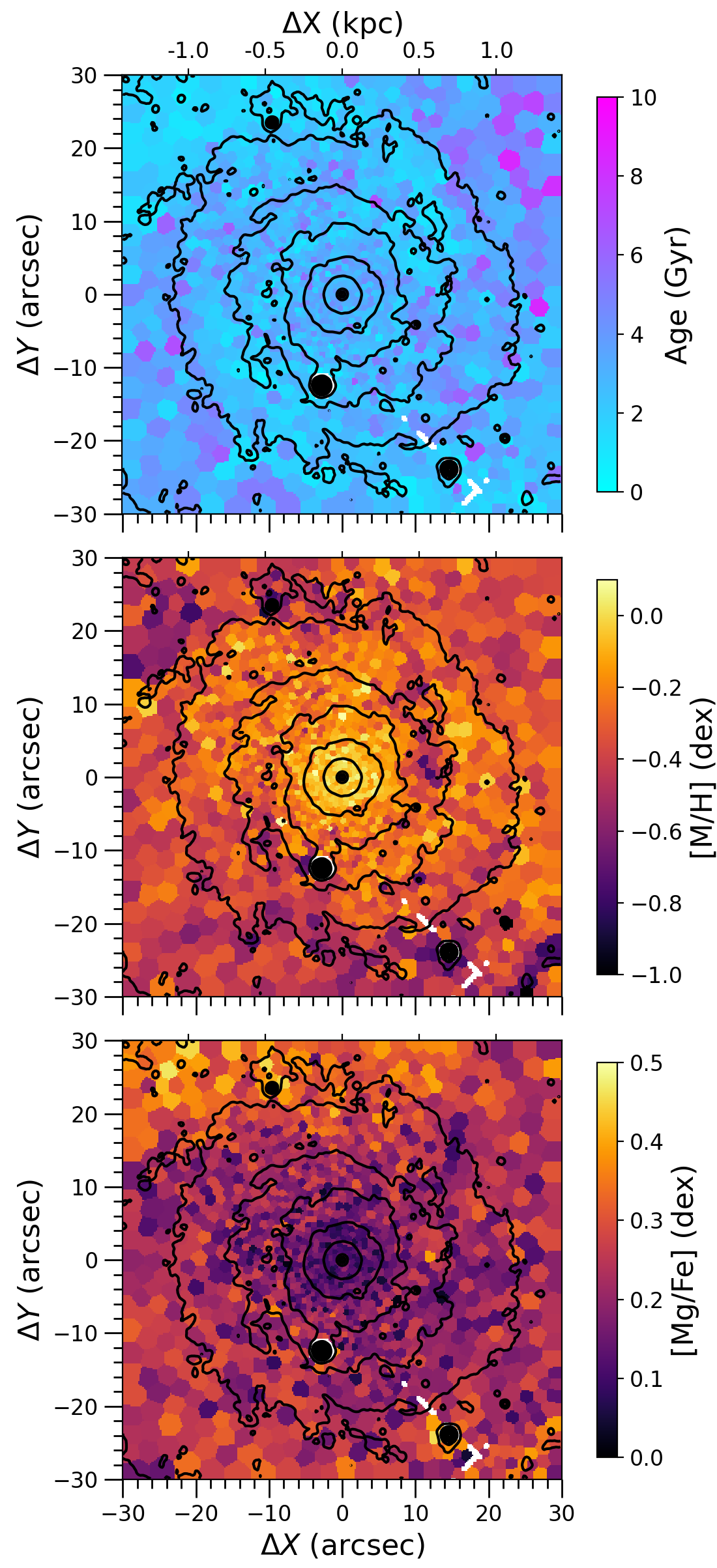}}
\caption{Stellar population maps of the central arcmin of M~74, extracted from the integrated (kinematic-corrected) data cube. From top to bottom: mean age, total metallicity [M/H] and [Mg/Fe] abundance. Masked regions are depicted in white and isophotal contours in black. The physical scale is given as a reference on the top $X$ axis. 
}
\label{fig:orig_pop_maps}
\end{figure}

We present here the results from the analysis of the kinematic-corrected original data cube of M~74 (Sect.~\ref{sub:preparation}). A Voronoi binning to a target S/N$=100$ was also used in this case. However, the different preprocessing compared to the host-galaxy data cube resulting from the C2D decomposition (Sect.~\ref{sub:c2d_m74}) leads to a different bin size compared to the host-galaxy data cube (Fig.~\ref{fig:host_pop_maps}). The analysis was performed following the method described in Sect.~\ref{sub:spectr_fit}. While the goal of the paper is to analyze the NSC and the host galaxy separately, the results for the integrated data cube provide a general context and serve as a reference. 
In Fig.~\ref{fig:orig_pop_maps}, we show the maps of the mean age, total metallicity ([M/H]) and [Mg/Fe] abundance, from top to bottom, extracted from the original MUSE data cube. 
Considering that Voronoi bins are, in general, larger in Fig.~\ref{fig:orig_pop_maps} than in Fig.~\ref{fig:host_pop_maps}, the maps in the two figures are totally compatible with each other. 
Similarly to Fig.~\ref{fig:host_pop_maps}, the age map (top panel of Fig.~\ref{fig:orig_pop_maps}) shows alternatively regions dominated by relatively young (younger than ${\sim}4$~Gyr) and old stars (older than ${\sim}8$~Gyr), roughly following the pattern of the spiral arms. 
[M/H] shows regions of higher and lower values. 
As in Fig.~\ref{fig:host_pop_maps}, there are regions with low metallicity in the outskirts of the mapped region, while we find supersolar metallicities within the central ${\sim}500$~pc. 
[Mg/Fe] abundances are in general anti-correlated with [M/H], with values relatively close to solar in the central ${\sim}500$~pc. 
Zooming into the very center of the middle panel in Fig.~\ref{fig:orig_pop_maps}, a clear [M/H] drop is seen in the PSF size. However, no significant age and [Mg/Fe] differences are seen in the very center of the top and bottom panels with this spatial binning level. 

\end{appendix}
\end{document}